\documentstyle[12pt,eqsecnum,aps,epsf]{revtex}
\tighten

\begin{document}

\def\lpmb#1{\mbox{\boldmath$#1$}}
\def\cd#1{{\cal D}#1}

\centerline{ \bf Numerical Approaches to High Energy Electroweak }
\centerline{ \bf Baryon Number Violation}
\centerline{ \bf Above and Below the Sphaleron Barrier
\footnote[1]{This paper  is based on talks given by 
  C. Rebbi and R. Singleton at the 1996 Summer School
  on Cosmology and High Energy Physics, held at  
  ICTP in Trieste, Italy.}
}

\vskip0.5cm
\centerline{Claudio Rebbi\footnote[2]{rebbi@pthind.bu.edu} }
\smallskip
\centerline{\it Physics Department}
\centerline{\it Boston University}
\centerline{\it 590 Commonwealth Avenue}
\centerline{\it Boston, MA 02215, USA}

\vskip0.7cm
\centerline{Robert Singleton, Jr.\footnote[3]
{bobs@terrapin.phys.washington.edu}}
\smallskip
\centerline{\it Physics Department}
\centerline{\it The University of Washington}
\centerline{\it Box 351560}
\centerline{\it Seattle, WA  98195, USA}

\setcounter{page}{0}
\thispagestyle{empty}

\vfill

\centerline{\bf Abstract}
\vskip0.2in
We review some promising numerical techniques 
for calculating high energy baryon number
violating cross sections in the standard model.
As these lectures are designed to be self-contained,
we present in some detail the formalism of 
Rubakov, Son, and Tinyakov, which provides a
means of bounding the two-particle cross sections
in a semi-classical manner. The saddle-point 
solutions required by this method must be found
computationally and are of two basic types,
corresponding to tunneling events between 
adjacent topological sectors on the one hand, 
and classically allowed evolution over the 
sphaleron barrier on the other. In both cases 
one looks for topology changing solutions of 
small incident particle number. In the 
classically allowed regime we have developed 
a Monte Carlo technique that systematically 
lowers the particle number while still ensuring 
that a change in topology takes place. We also 
make progress towards a numerical method 
amenable for the more computationally challenging 
problem of finding the complexified tunneling 
solutions, and we present some of our numerical
findings, both above and below the  sphaleron 
barrier.

\bigskip

\noindent BUHEP-96-36 \hskip 0.5cm  UW/PT 96-21

\noindent hep-ph/9706424 \hfill Typeset in La\TeX

\eject	

\baselineskip 24pt plus 2pt minus 2pt

\section{Introduction}
\hspace{1cm}

The prospect\cite{ring,esp} of observable baryon number
violation in  high energy electroweak collisions has
provoked much excitement, but despite considerable
effort, it has not been possible to obtain conclusive
evidence that unsuppressed baryon number violation can
indeed occur at energies in the multi-TeV range.
The purpose of this paper is to summarize some recent 
developments which offer hope for a reliable calculation 
of the magnitude of baryon number violating cross sections.

Limitations of space do not permit a comprehensive
treatment of this vast subject.  These lectures will
focus on the semiclassical methods introduced in
Refs.~\cite{rs} and \cite{rst} and further work by the 
present authors. The main idea in Refs.~\cite{rs} and 
\cite{rst} consists in studying processes with energy
and initial particle number of the form $E=\epsilon / g^2$ 
and $N_i = \nu /g^2$, with $g$ being the electroweak 
coupling constant, in the limit where $g \to 0$ while 
$\epsilon$ and $\nu$ are held fixed.  Under such 
conditions one can justify the use of semiclassical 
methods to extract the exponential behavior of the 
semi-exclusive cross section, which can be done by
calculating hybrid Euclidean-Minkowski solutions which
violate baryon number via tunneling from one
(non-vacuum) real-time configuration to another.
Finding such such semiclassical solutions
is however highly non-trivial and can only be
done by computational methods. In Ref.~\cite{rs} 
we have applied numerical techniques to the study 
of processes in which the gauge and Higgs fields 
change their topology through purely classical 
evolution, thus inducing a violation of  baryon 
number. The investigation of the semiclassical 
solutions that account for tunneling under the 
barrier is in progress.

These two approaches, one involving classically
forbidden tunneling-like processes and the other
classically allowed topological transitions, probe
complementary aspects of the problem and should
produce compatible results (thereby providing
consistency checks). Both approaches require
solutions to certain nonlinear partial
differential equations for which no (nontrivial)
exact solutions are known; however, these
equations are well suited to computational
study and one can still make considerable
progress. Along with P. Tinyakov, we have
recently launched a numerical investigation
of the standard model baryon number violating
rates based on Ref.~\cite{rst}, and in the next
section we shall review the relevant formalism
and present some of our initial results. The
rest of this paper is devoted to the numerical
work of Ref.~\cite{rs}. We presently have much
more to say about this approach because
its corresponding computational study is at a
mature stage of development. Finally, in an
effort to write a self contained work, the
remainder of this introduction is devoted to
a brief exposition of nonperturbative baryon
number violation in the standard model of
electroweak interactions.

For our purposes, when we talk of the
``standard model'' we mean the standard model 
in which the Weinberg angle has been set to zero,
i.e. we shall be considering $SU(2)$ gauge
theory spontaneously broken via a single Higgs 
doublet. This simplified model has all the
relevant physics. Most important, the 
gauge structure dictates nontrivial topology 
for the bosonic vacuum sector. Working in
the temporal gauge with periodic boundary
conditions at spatial infinity, each
vacuum may be characterized by
an integer called the winding number which
measures the number of times the gauge
manifold is wound around 3-space\cite{JR}. 
As this number is a topological invariant, 
vacua of different winding numbers 
cannot be continuously deformed into 
one another.  

Because of the axial vector anomaly, baryon number
violation occurs when the gauge and Higgs fields
change their topology\cite{thooft76}. Adjacent
topological sectors are separated by an extremely 
high barrier, the top of which is a static saddle-point 
solution to the equations of motion. This configuration 
is called the sphaleron\cite{KM}, and it has an energy
of about $10 \, {\rm TeV}$ and possesses a single 
unstable direction in field space. At low energy the 
baryon number violating rates are exceedingly small, 
as the gauge and Higgs fields must  first tunnel 
through the sphaleron, which is indeed extremely 
unlikely. 

Recently, the prospect of rapid baryon number
violation at high temperatures and high energies 
has emerged. The basic idea is that if the gauge
and Higgs fields have enough energy, they can
change their topology by sailing over the sphaleron 
barrier rather than being forced to tunnel through 
it. At high temperatures this is precisely what
happens, and it is generally agreed that baryon
number violation becomes unsuppressed in the
early universe\cite{highT}. 

The situation in high energy collisions is far 
less clear. The limiting process in baryon 
number violation is the production of a 
sphaleron-like configuration from an
incident beam of high energy particles. 
But since the sphaleron is a large extended 
object, there is a scale mismatch with the 
initial high energy two-particle state, and 
hence one naively expects the baryon number 
violating rate to be small. However, Ringwald
\cite{ring} and Espinosa\cite{esp} have suggested
that the sum over many-particle final states gives
rise to factors that grow with energy sufficiently
rapidly to offset any exponential suppression.
If true, this offers the exciting prospect of 
one day observing baryon number violation in
high energy collisions.

\section{The Classically Forbidden Domain}
\hspace{1cm}

The approach of Ringwald and Espinosa\cite{ring,esp}, 
however, neglects some important corrections which still 
elude calculation despite considerable effort. Apart from 
lattice simulations, semi-classical techniques are our only 
handle on nonperturbative effects. The basic problem 
with calculating the anomalous baryon number violating 
cross sections is that exclusive two-particle initial states 
are not very amenable to these methods, and there are 
potentially large initial state corrections whose effects 
remain undetermined.  

Rather than calculating the two-particle cross section
directly, Rubakov, Son and Tinyakov\cite{rst} 
investigate a related quantity for which semiclassical
methods are still applicable. Their method involves 
saturating the path integral representation of this 
quantity with a complexified Euclidean-Minkowski 
saddle-point. This solution includes the effects of 
tunneling under the sphaleron barrier, and is a 
generalization of the periodic instanton of Ref.~\cite{krt}. 
We now review in more detail the work of Ref.~\cite{rst}, 
along with selected portions of Refs.~\cite{krt} --\cite{rt}  
upon which this work is based.

\subsection{The Inclusive Cross Section}

As previously mentioned, the calculations of 
Refs.~\cite{ring} and \cite{esp} for the two-particle
baryon number violating cross section, $\sigma_2(E)$,  
become unreliable at high energy. This is because 
of a failure of semiclassical methods in calculating
exclusive quantities like two-particle scattering
amplitudes. So rather than calculating $\sigma_2(E)$
directly, Ref.~\cite{rst}  examines a related {\it inclusive} 
quantity:
\begin{eqnarray}
\label{sigincl}
  \sigma(E,N) &=&   \sum_{f,i} \, 
  \mid<f| \, \hat S \, \hat P_E\, \hat P_N \, |i>\mid^2 \ ,
\end{eqnarray}
where the sum is over all initial and final states, 
$\hat S$ is the $S$-matrix, and $\hat P_E$ and 
$\hat P_N$ are projection operators onto subspaces 
of energy $E$ and particle number $N$ respectively.  

Unlike the exclusive two-particle cross section, 
$\sigma(E,N)$ is directly calculable by semiclassical 
methods as long as the incident particle number $N$ 
remains large. If the energy and particle number are
parameterized by
\begin{mathletters}%
\label{ENgsp}
\begin{eqnarray}
\label{Egsq} 
  E &=& {\epsilon \over g^2} 
\\
\label{Ngsq}
  N &=& {\nu \over g^2} \ ,
\end{eqnarray}
\end{mathletters}%
and the fields are rescaled by appropriate powers
of $g^2$, then in the limit $g \to 0$ with $\epsilon$ 
and $\nu$ held fixed, the path integral for $\sigma(E,N)$ 
can be saturated by a  $g$-independent classical 
saddle-point solution to the equations of motion. As 
shown in the next section, the cross section takes the 
form
\begin{eqnarray}
\label{sigsc}
  \sigma(E,N) &=&  
  \exp \left[ {1 \over g^2} \, 
  F(\epsilon,\nu) + {\cal O}(g^0)\right]  \ ,
\end{eqnarray}
where the function $F(\epsilon,\nu)$ is determined
by the classical solution. 


The utility of $\sigma(E,N)$ is that it may be used
to bound the two-particle cross section and
allow one to extract the exponential behavior
of $\sigma_2(E)$. By construction, $\sigma(E,N)$
provides an upper bound to $\sigma_2(E)$. 
This is because one of the initial $N$-particle states 
of  (\ref{sigincl}) possesses $N-2$ free propagating
particles and two colliding particles\cite{pt}. 
A lower bound may be obtained under some reasonable 
physical assumptions~\cite{modproj}.  Let $|\psi_N>$ 
be the initial state that saturates the sum in (\ref{sigincl}).  
If the process \hbox{$2\to any$} proceeds through some 
preferred intermediate state, such as a sphaleron-like 
configuration in the case of  baryon number violation, 
then the substitution of this state by $|\psi_N>$ will 
underestimate the result, giving the inequality 
$\mid <\psi_N|2> \mid^2 \sigma(E,N) < \sigma_2(E)$.
Estimating $\mid <\psi_N|2> \mid^2 \sim \exp(-const \, N)$,
together with the previous upper bound, gives the 
inequalities
\begin{eqnarray}
\label{sigineq}
  \exp(-const \, N) \, \sigma(E,N) < \sigma_2(E) \,
  < \sigma(E,N) \ ,
\end{eqnarray}
{}from which it follows that
\begin{eqnarray}
  \lim_{g \to 0}  g^2 \ln \sigma_2(E) = F(\epsilon,\nu) + \cal{O(\nu)} \ .
\end{eqnarray}
The consistency of the first inequality requires
that $F(\epsilon,\nu)$ has a smooth $\nu \to 0$ 
limit, in which case $F(\epsilon,0)$ determines 
the exponential behavior of $\sigma_2(E)$. 
However, the second inequality of (\ref{sigineq}) 
holds regardless of continuity, and hence
if $\sigma(E,N)$ is exponentially suppressed 
(for any value of $N$), then so is $\sigma_2(E)$. 

\subsection{Development of the Formalism}

We now review the formalism developed in Ref.~\cite{rst}
used to calculate the inclusive-state cross section
$\sigma(E,N)$. For purposes of illustration, we consider 
a system with a single real scalar field, whose field 
operator $\hat\phi({\bf x})$ has eigenstates defined by 
\hbox{$\hat\phi({\bf x})\,|\phi>\,=\phi({\bf x})\,|\phi>$}. 
The approach that follows is based on a coherent state 
formalism, where coherent states $|a>$ are defined by 
$\hat a_{\bf k}\,|a>\,=a_{\bf k}\,|a>$, with $\hat a_{\bf k}$ 
being the annihilation operator of the ${\bf k}$-th mode. 
In field space the coherent states take the form
\begin{eqnarray}
\label{fieldrepphia}
  <\phi | a> \, = {\rm const} \cdot \exp\bigg[
  \int d^3k \left\{
  -\frac{1}{2} \, a_{\bf k} a_{-\bf k} - \frac{1}{2} \,
  \omega_{\bf k} \phi({\bf k})\phi(-{\bf k}) 
  +  \, \sqrt{2\omega_{\bf k}} \, a_{\bf k} \phi({\bf k}) 
  \right\}\bigg] \ ,
\end{eqnarray}
where $\phi({\bf k})$ is the spatial Fourier transform
of $\phi({\bf x})$, given by
\begin{eqnarray}
  \phi({\bf x}) =   \int \frac{d^3k}{(2\pi)^{3/2}} \,
  e^{i {\bf k}\cdot {\bf x}} \, \phi({\bf k}) \ .
\end{eqnarray}

In the coherent state formalism, the $S$-matrix  is 
represented by its kernel $S(b^*,a) \equiv <b|\hat S|a>$, 
and inserting a complete set of field-states we
can write
\begin{eqnarray}
\label{srep}
  S(b^*,a) = \int \cd{\phi_i({\bf x}) }\cd{\phi_f({\bf x})} \,
  <b|\phi_f><\phi_f|\hat S|\phi_i><\phi_i|a> \ .
\end{eqnarray}
Upon explicitly extracting the time dependence from the 
annihilation operators in (\ref{srep}) and (\ref{fieldrepphia})  
by writing $a_{\bf k} \to a_{\bf k}  \, e^{-i\omega_{\bf k}T_i}$ 
and $b^*_{\bf k} \to b^*_{\bf k}  \, e^{i\omega_{\bf k}T_f}$, 
and using the functional integral representation of 
$<\phi_f|\hat S|\phi_i>$, one can write
\begin{eqnarray}
\label{kerns}
  S(b^*,a) = \int  \cd{\phi(x)} \, \exp\bigg[
  i S[\phi] + B_i(\phi_i,a) + B_f(\phi_f,b^*)\bigg] \ ,
\end{eqnarray}
\noindent
where the integral over the boundary configurations
$\phi_i$ and $\phi_f$ and the integral over all paths
interpolating between these configurations have been
combined into a single path integral, and the
boundary terms are given by
\begin{mathletters}%
\label{bifdef}
\begin{eqnarray}
  B_i(\phi_i,a) &=& \int d^3 k \left\{
  -\frac{1}{2} \, a_{\bf k} a_{-\bf k}\, 
  e^{-2i\omega_{\bf k}T_i} - \frac{1}{2} \,\omega_{\bf k}
  \phi_i({\bf k})\phi_i(-{\bf k}) + \sqrt{2\omega_{\bf k}} \, 
  a_{\bf k} \, e^{-i\omega_{\bf k}T_i}  
  \phi_i({\bf k}) \right\} 
\\ \nonumber \\ 
  B_f(\phi_f,b^*) &=& \int d^3 k \left\{
  -\frac{1}{2} \, b^*_{\bf k} b^*_{-\bf k} \, 
  e^{2i\omega_{\bf k}T_f} - \frac{1}{2} \,\omega_{\bf k} 
  \phi_f({\bf k})\phi_f(-{\bf k}) +  \sqrt{2\omega_{\bf k}} \, 
  b^*_{\bf k}\, e^{i\omega_{\bf k}T_f} 
  \phi_f(-{\bf k}) \right\}\ .
\end{eqnarray}
\end{mathletters}%

To proceed, the sums over initial and final states 
in (\ref{sigincl}) are replaced by
\begin{mathletters}%
\begin{eqnarray}
  \sum_i &&\to \int \cd{a^*_{\bf k}}\cd{a_{\bf k}} \, 
  \exp \left[-\int d^3k \, a^*_{\bf k} a_{\bf k}  \right]
\\ 
  \sum_f &&\to \int \cd{b^*_{\bf k}}\cd{b_{\bf k}} \, 
  \exp \left[-\int d^3k \, b^*_{\bf k} b_{\bf k}  \right] \ ,
\end{eqnarray}
\end{mathletters}%
and unity, in the form  
\begin{eqnarray} 
  \int \cd{c_{\bf k}^*} \, \cd{c_{\bf k}} \,
  \exp\left[-\int d^3k c_{\bf k}^*c_{\bf k}
   \right] \, | c><c| = {\bf 1} \ ,
\end{eqnarray}
is inserted between $\hat S$ and $\hat P_E\hat P_N$, 
giving
\begin{eqnarray} 
\label{siginter}
  \sigma(E,N) &=& \int {\cal D}[a,b,c,e] \, \exp\left[
  -b^*b - a^*a - c^*c - e^*e \right]
\\ \nonumber && 
  S(b^*,c) S(b^*,e)^* <c\mid\hat P_E  \hat P_N\mid a > \,
   <a\mid\hat P_E \hat P_N\mid e > \ .
\end{eqnarray}
We are using an obvious short-hand notation for the
integration measure, and integrals over momenta
are implied.  The kernels of the projection operators 
take the form
\begin{mathletters}%
\begin{eqnarray}
 <b | \hat P_E|a> &=& \int d \xi \, 
  \exp\left[-i E \xi + \int d^3k \, 
  e^{i \omega_{\bf k}\xi}\,b^*_{\bf k}a_{\bf k}  
  \right] 
\\
 <b | \hat P_N|a> &=& \int d \eta \, 
  \exp\left[-i N \eta + \int d^3k \, 
  e^{i \eta}\,b^*_{\bf k}a_{\bf k}  
  \right] \ ,
\end{eqnarray}
\end{mathletters}%
{}from which it follows that 
\begin{eqnarray}
\label{kernpepn}
 <b | \hat P_E \hat P_N|a> &=& \int d\xi d\eta\, 
  \exp\left[-i E \xi - iN \eta+ \int d^3k \, 
  e^{i \omega_{\bf k}\xi +i \eta}\,b^*_{\bf k}
  a_{\bf k}  \right]  \ .
\end{eqnarray}

After  substituting  (\ref{kerns}) and (\ref{kernpepn}) 
into (\ref{siginter}), and then changing variables 
via $a\to \exp[-i\omega \xi -i \eta] \, a$ 
and $a^*\to \exp[-i\omega \xi' -i \eta'] \, a^*$, 
the $c$-integral may be performed to obtain
\begin{eqnarray} 
  \int \cd{c^*}\cd{c} \, \exp\bigg[- c^* c + c^* a 
  + B_i(\phi_i,c) \bigg] = \exp[B_i(\phi_i,a)] \ ,
\end{eqnarray}
with a similar expression for the $e$-integration. 
Finally, after collecting terms and redefining 
\hbox{$\xi+\xi' \to \xi$} and \hbox{$\eta+\eta' \to \eta$}, 
the cross section becomes
\begin{mathletters}%
\label{sigw}
\begin{eqnarray}
  \sigma(E,N) = \int \cd{\phi(x)}\cd{\phi'(x)}
  \cd{a_{\bf k}^*} \cd{a_{\bf k}} \cd{b_{\bf k}^*} 
  \cd{b_{\bf k}} \, d\eta  d\xi \, \exp\left[ W \right] \ ,
\end{eqnarray}
where
\begin{eqnarray}
\label{Wdef}
  W &=&  -i E \xi -i N \eta - \int d^3k \left\{ 
  b^*_{\bf k} b_{\bf k} + a^*_{\bf k} a_{\bf k} \,
  e^{-i \Delta_{\bf k}} \right\}
\\ \nonumber
  &+& \, i S[\phi] - i S[\phi'] + B_i(\phi_i,a) + B_f(\phi_f,b^*) 
  + B_i(\phi_i',a)^* + B_f(\phi_f',b^*)^* \ ,
\end{eqnarray}
\end{mathletters}%
with $\Delta_{\bf k}=\omega_{\bf k} \xi + \eta$.
The functional $S[\phi(x)]$ is the action, and 
the boundary terms at the initial and final times 
$T_i$ and $T_f$ are given by (\ref{bifdef}).

To display the semiclassical nature of the cross 
section, it is convenient to express the exponential
factor $W$ in terms of the rescaled field $\tilde\phi=
g\,\phi$, the rescaled mode amplitudes $\tilde a_{\bf k}=
g \, a_{\bf k}$ and $\tilde b_{\bf k}=g \, b_{\bf k}$, 
and the rescaled energy $\epsilon$ and particle number 
$\nu$ defined in (\ref{ENgsp}). The action $\tilde 
S[\tilde \phi]$, which is related to the unscaled 
action by \hbox{$S[\phi]=\tilde S[\tilde\phi]/g^2$},
is $g$-independent and we can thus write $W(E,N)=
F(\epsilon,\nu)/g^2$, with the function $F$ being 
independent of the coupling constant. Hence, for
small values of $g$, it is a good approximation to 
simply saturate the integrals by classical saddle-point 
solutions, from which we obtain 
\begin{eqnarray}
\label{sigf}
  \sigma(E,N) &=&  
  \exp \left[ {1 \over g^2} \, 
  F(\epsilon,\nu)\right]  \ ,
\end{eqnarray}
where $F$ is determined by evaluating (\ref{Wdef}) 
on the classical solution. In what follows we shall
work only with the rescaled quantities in which the
$g$-dependence has been factored out, but for notational 
simplicity we will use the unscaled notation and drop 
the tilde over the associated quantity. 

In looking for saddle-points of (\ref{Wdef}) we must 
distinguish between two cases. There may be solutions 
which correspond to classically allowed evolution, in 
which case the fields and the action will be real and 
the parameter $\Delta_{\bf k}$ zero. As shown below, 
this implies that the function $F$ of (\ref{sigf}) 
will be zero, signalling the absence of suppression, 
and if we can find such classical solutions with small 
$\nu$, this furthermore indicates that the two-particle 
rates are likewise unsuppressed. Classically allowed 
evolution which changes the topology of the fields must 
perforce occur at an energy above the sphaleron barrier, 
but $E>E_{\rm sph}$ is per se not a sufficient condition 
for the existence of classically allowed solutions with 
a given particle number in the initial state, and in a
later section we shall return to the problem of finding 
topology changing solutions with low incident particle 
number.

Alternatively, there may be solutions which correspond 
to classically forbidden processes, in which case the
saddle-points for $\phi$ and $\phi'$, \hbox{$\delta 
S[\phi]/\delta \phi=0$} and \hbox{$\delta S[\phi']/ 
\delta \phi'=0$}, may in fact have imaginary components,
while the saddle-point values of $a$ and $a^*$ need
not be complex conjugates. Obtaining these complexified 
saddle-points is much more involved than finding 
classically allowed solutions passing over the 
sphaleron barrier, and so we devote the remainder 
of this section to explicating some of the details 
of the procedure, with a special emphasis on boundary 
conditions. 

Extremizing (\ref{Wdef}) with respect to
modes $a_{\bf k}$ and $a^*_{\bf k}$ yields
\begin{mathletters}
\label{dela}
\begin{eqnarray}
\label{delaa}
   a_{\bf k}^*\, e^{-i \Delta_{\bf k}} + a_{-\bf k}  
  \, e^{-2 i  \omega T_i} -
  \sqrt{2\omega} \,  \phi_i({\bf k}) \, 
  e^{- i \omega T_i} &=& 0
\\
\label{delab}
   a_{\bf k}\, e^{-i \Delta_{\bf k}} + a_{-\bf k}^*  \, 
  e^{2 i \omega T_i} - \sqrt{2\omega} \,  \phi_i'({-\bf k}) 
  \, e^{i  \omega T_i} &=& 0
\ ,
\end{eqnarray}
\end{mathletters}
which may be solved to give the saddle-points
\begin{mathletters}%
\label{asaddle}
\begin{eqnarray}
  a_{\bf k} &=& \frac{\sqrt{2 \omega_{\bf k}} }
  {e^{-i\Delta_{\bf k}} - 
  e^{i\Delta_{\bf k}}} \, \left[ \phi_i'(-{\bf k})
  -e^{i\Delta_{\bf k}} \phi_i(-{\bf k})  \right] \, 
  e^{i\omega_{\bf k} T_i}
\\
  \bar a_{\bf k}  &=& \frac{\sqrt{2 \omega_{\bf k}} }
  {e^{-i\Delta_{\bf k}} - e^{i\Delta_{\bf k}}} \, 
  \left[ \phi_i({\bf k})-e^{i\Delta_{\bf k}} 
  \phi_i'({\bf k})\right] \, e^{-i\omega_{\bf k} T_i} \ .
\end{eqnarray}
\end{mathletters}%
As previously noted, in general these solutions are not
complex conjugates, and hence we use the bar notation 
for the latter. The expression (\ref{asaddle}) relates 
the $a$-mode amplitudes to the initial saddle-point
values of the incident fields, which in turn are 
constrained by
\begin{mathletters}
\label{delphii}
\begin{eqnarray}
\label{delphiia}
  -i \dot\phi_i({\bf k}) - \omega \phi_i({\bf k}) + 
  \sqrt{2\omega} \, a_{-\bf k} \, e^{- i  \omega T_i} &=& 0
\\
\label{delphiib}
  i \dot\phi_i'({\bf k}) - \omega \phi_i'({\bf k}) + 
  \sqrt{2\omega} \,  a_{\bf k}^* \, e^{ i  \omega T_i} &=& 0 \ ,
\end{eqnarray}
\end{mathletters}
and with the use of (\ref{asaddle}), one can write this
expression as a boundary condition involving only the 
incident fields,
\begin{mathletters}%
\label{phiisaddle}
\begin{eqnarray}
  i \dot \phi_i({\bf k}) + \omega \, \phi_i({\bf k})&=&
  e^{i\Delta_{\bf k}} \, \left[ i \dot \phi_i'({\bf k}) + 
  \omega \, \phi_i'({\bf k}) \right]
\\ 
  i \dot \phi_i({\bf k}) - \omega \, \phi_i({\bf k})&=&
  e^{-i\Delta_{\bf k}} \, \left[ i \dot \phi_i'({\bf k}) 
  - \omega \, \phi_i'({\bf k}) \right] \ .
\end{eqnarray}
\end{mathletters}%
The parameter $\Delta_{\bf k}$ itself is determined
by saddle-point equations, and when it vanishes
note that $\phi'_i({\bf k})=\phi_i({\bf k})$, and 
that $a_{\bf k}=(\omega_{\bf k}/2)^{1/2}\,
\phi_i(-{\bf k}) \, e^{i\omega_{\bf k} T_i}$ and 
$\bar a_{\bf k}=(\omega_{\bf k}/2)^{1/2}\,
\phi_i({\bf k}) \,e^{-i\omega_{\bf k} T_i}$ are 
complex conjugates. This case therefore corresponds 
to a classically allowed process above the 
sphaleron barrier. 

To obtain final-state boundary conditions, one extremes
(\ref{Wdef}) with respect to the $b$-modes and the final-state 
fields $\phi_f$ and $\phi'_f$, which gives
\begin{mathletters}
\label{delb}
\begin{eqnarray}
\label{delba}
   b_{\bf k}^* + b_{-\bf k}  \, e^{-2 i  \omega T_f} -
  \sqrt{2\omega} \,  \phi_f'({\bf k}) \, 
  e^{- i  \omega T_f} &=& 0 
\\ 
\label{delbb}
   b_{\bf k} + b_{-\bf k}^*  \, e^{2 i  \omega T_f} -
  \sqrt{2\omega} \,  \phi_f({-\bf k}) \, 
  e^{i  \omega T_f} &=& 0 \ ,
\end{eqnarray}
\end{mathletters}
\noindent
and
\begin{mathletters}
\label{delphif}
\begin{eqnarray}
\label{delphifa}
  i \dot\phi_f({\bf k}) - \omega \phi_f({\bf k}) + 
  \sqrt{2\omega} \, b_{\bf k}^* \, e^{i  \omega T_f} &=& 0
\\
\label{delphifb}
  -i \dot\phi_f'({\bf k}) - \omega \phi_f'({\bf k}) + 
  \sqrt{2\omega} \,  b_{-\bf k} \, e^{ -i  \omega T_f} &=& 0 \ .
\end{eqnarray}
\end{mathletters}
Together, (\ref{delb}) and (\ref{delphif}) imply 
that the final-state fields and their respective time 
derivatives agree, \hbox{$\phi_f'({\bf k}) =\phi_f({\bf k})$} 
and \hbox{$\dot\phi_f'({\bf k})= \dot\phi_f({\bf k})$}.
Thus, as the saddle-points $\phi$ and $\phi'$ satisfy 
the same classical equations at intermediate times, 
they also agree at these times, and hence we are really 
dealing with a single solution $\phi(x)$. At first 
sight this seems inconsistent with (\ref{phiisaddle}) 
for nonzero $\Delta_{\bf k}$. However, the general 
complex saddle-point solution is nonanalytic, and 
$\phi_i$ and $\phi_i'$ in (\ref{phiisaddle}) are to 
be thought of as lying on separate sheets in the 
complex-$t$ plane (to emphasize this, we will not 
remove the prime from $\phi_i'$).

The value of $\Delta_{\bf k}$ is determined by
the saddle-points of $\xi$ and $\eta$, which from 
(\ref{Wdef}) are related to the energy and particle 
number by 
\begin{mathletters}%
\label{delen}
\begin{eqnarray}
\label{delena}
  \epsilon &=&  \int d^3k \,  \omega_{\bf k} \, 
  a_{\bf k}^* a_{\bf k} \,e^{-i\Delta_{\bf k}}
\\ 
\label{delenb}
  \nu &=& \int d^3k  \, a_{\bf k}^* 
  a_{\bf k} \, e^{-i\Delta_{\bf k}} \ .
\end{eqnarray}
\end{mathletters}%
The saddle-point of $\xi$ may be made pure imaginary
by a suitable time-translation, and the real part of the
$\eta$-saddle-point is typically small\cite{pt}, so we 
can write
\begin{mathletters}%
\begin{eqnarray}
  \xi &=& i T
\\ 
  \eta &=& i \theta \ .
\end{eqnarray}
\end{mathletters}%
The parameter $T$ can be removed from the boundary
conditions by choosing the complex time contours
of Fig.~1.  Since the fields become linear in the 
distant past we can write
\begin{figure}
\centerline{
\epsfxsize=90mm
\epsfbox[0 0 640 427]{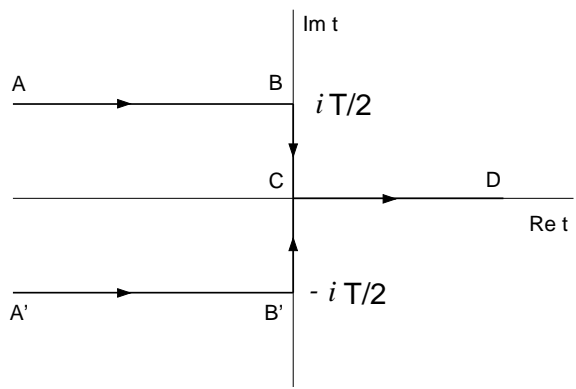}
}
\caption{\tenrm
Complex-time contours
}
\end{figure}
\begin{mathletters}%
\label{philin}
\begin{eqnarray}
\label{philina}
  \phi({\bf k}) &=& \frac{1}{\sqrt{2\omega_{\bf k}}}
  \left[ f_{\bf k} \,e^{-i\omega_{\bf k} \tau} 
  + g_{\bf k} \,e^{i\omega_{\bf k} \tau} \right]
   \qquad{\rm on\  line\  AB}
\\ 
\label{philinb}
  \phi'({\bf k}) &=&  \frac{1}{\sqrt{2\omega_{\bf k}}}
  \left[f'_{\bf k} \,e^{-i\omega_{\bf k}\tau} 
  + g'_{\bf k} \,e^{i\omega_{\bf k} \tau} \right]
  \qquad{\rm on\  line\  A'B'} 
\end{eqnarray}
\end{mathletters}%
as $\tau ={\rm Re}\,t \to -\infty$.
The boundary conditions (\ref{phiisaddle}) will be applied along
line $AB$, keeping in mind that $\phi_i$ and $\phi_i'$ lie on  
different sheets in this region. In this asymptotic linear
domain, however, the fields are analytic on their respective 
sheets, and hence
\begin{eqnarray}
\label{philincont}
  \phi'({\bf k}) &=&  \frac{1}{\sqrt{2\omega_{\bf k}}}
  \left[ f'_{\bf k} \,e^{\omega T -i\omega_{\bf k}\tau} 
  + g'_{\bf k} \,e^{-\omega T +i\omega_{\bf k} \tau} \right]
   \qquad{\rm on\  line\  AB} 
\end{eqnarray}
as $\tau \to -\infty$. 
This relation, with (\ref{philina}) and boundary
conditions (\ref{phiisaddle}), gives the restriction
\begin{mathletters}%
\label{fg}
\begin{eqnarray}
  f'_{\bf k} &=& e^\theta \, f_{\bf k} 
\\ 
  g_{\bf k}' &=& e^{-\theta} \, g_{\bf k}  \ ,
\end{eqnarray}
\end{mathletters}%
and the degrees of freedom associated with the field 
along $A'B'$ have been eliminated.

A number of simplifications now occur. 
The energy and particle number may be written
\begin{mathletters}%
\label{delenfg}
\begin{eqnarray}
\label{delenfga}
  \epsilon &=&  \int d^3k \,  \omega_{\bf k} \, 
  f_{\bf k}^* \,g_{-\bf k}
\\ 
\label{delenfgb}
  \nu &=& \int d^3k  \, f_{\bf k}^* \,g_{-\bf k} \ .
\end{eqnarray}
\end{mathletters}%
Upon taking the limits $\,T_i \to -\infty$ and 
$T_f \to \infty$ the boundary terms become
\begin{mathletters}%
\begin{eqnarray}
  B_i(\phi_i,a) &=& B_i^*(\phi_i',a) \hskip0.2cm = 
  \frac{1}{2}\int d^3k \, f_{-\bf k} \, g_{\bf k}
\\ 
  B_f(\phi_f,b^*) &=& B_f^*(\phi_f',b^*) = 
  \frac{1}{2}\int d^3k \, b^*_{\bf k} \, b_{\bf k} \ ,
\end{eqnarray}
\end{mathletters}%
and thus the exponential factor on the solution takes 
the form
\begin{eqnarray}
\label{wsaddle}
  F = \epsilon \, T + \nu \theta + i S[\phi] - i S[\phi'] \ ,
\end{eqnarray}
where the (rescaled $g$-independent) actions $S[\phi]$ 
and $S[\phi']$ are evaluated along the upper contour 
$ABCD$ on the first and second sheets respectively. 

By virtue of a symmetry akin to $CPT$, these  
expressions simplify considerably if the saddle-point  
is unique. Since the coefficients of the field equations 
$\delta S[\phi]/\delta \phi =0$ are real, given a 
solution $\phi({\bf x},t)$, one can form a new solution 
$\Phi({\bf x},t) = \phi({\bf x},t^*)^*$. Uniqueness 
then implies the conjugation symmetry $\phi({\bf x},t)
=\phi({\bf x},t^*)^*$, and hence $f'_{\bf k} = 
g^*_{-\bf k}$ and $g'_{\bf k}=f^*_{-\bf k}$. This
can be used to express (\ref{fg}) as
\begin{eqnarray}
\label{gfnew}
  g_{\bf k}=e^\theta \, f_{-\bf k}^* \ ,
\end{eqnarray}
{}from which it follows that the energy and particle
number take the form
\begin{mathletters}%
\label{delenth}
\begin{eqnarray}
\label{delentha}
  \epsilon &=&  e^{-\theta}\,\int d^3k \,  
  \omega_{\bf k} \, f_{\bf k}^* f_{\bf k}
\\ 
\label{delenthb}
  \nu &=& e^{-\theta}\,\int d^3k  \, 
  f_{\bf k}^* f_{\bf k} \ .
\end{eqnarray}
\end{mathletters}%

Expression (\ref{gfnew}) may also be used 
to rewrite the boundary conditions (\ref{philin}) 
in the rather convenient form
\begin{mathletters}%
\label{philinun}
\begin{eqnarray}
\label{philinuna}
  \phi({\bf k}) &=& \frac{1}{\sqrt{2\omega_{\bf k}}}
  \left[ f_{\bf k} \,e^{-i\omega_{\bf k} \tau} 
  + e^\theta \, f_{\bf k}^* \,e^{i\omega_{\bf k} \tau} \right]
   \qquad{\rm on\  line\  AB}
\\ 
\label{philinunb}
  \phi'({\bf k}) &=&  \frac{1}{\sqrt{2\omega_{\bf k}}}
  \left[e^\theta \, f_{\bf k} \,e^{-i\omega_{\bf k}\tau} 
  + f_{\bf k}^* \,e^{i\omega_{\bf k} \tau} \right]
  \qquad{\rm on\  line\  A'B'} 
\end{eqnarray}
\end{mathletters}%
as $\tau \to -\infty$. Note that the conjugation 
symmetry implies that the solution is real along 
the entire real-Minkowski axis, and we shall thus 
impose the additional boundary condition ${\rm Im}\,
\phi({\bf x},t=0)=0$. In general, however, the 
solution becomes complex along the time-contours 
$ABC$ and $A'B'C$, and the consistency of 
(\ref{philinuna}) and (\ref{philinunb}) requires 
that $\phi(x)$ also possesses singularities between these 
contours. For the case in which $\theta=0$, the
field becomes real along $AB$ and $A'B'$ in the 
infinite past, and hence it remains real along 
the entire upper and lower contours. When the 
solution is also real along the imaginary-time 
axis, as is the case for periodic instantons\cite{krt}, 
turning-point boundary conditions $\dot \phi=0$
are also satisfied at $B$, $B'$ and $C$.

Recall that in  (\ref{wsaddle}), both $S=S[\phi]$ and 
$S'=S[\phi']$ are evaluated along the contour $ABCD$, 
albeit on different sheets in the complex-$t$ plane. 
If the singularities of $\phi$ only lie between $AB$ 
and $A'B'$,  the action $S'$ along $ABCD$ is equal 
to the action along $A'B'CD$ (staying on the same 
sheet, and assuming no contribution from the contour 
at infinity). Hence, the conjugation symmetry implies 
$S'=S^*$, and thus
\begin{eqnarray}
  F = \epsilon\, T + \nu \theta  - 2\, {\rm Im} 
  \, S(T,\theta) \ ,
\end{eqnarray}
where the implicit $T$ and $\theta$ dependence in 
the action has been made explicit. Note that 
$S(T,\theta)$ can be obtained by integrating
only along $ABC$, as the contribution from the
Minkowski section $CD$ is real and does not
contribute to $F$. Furthermore, since $T$ and 
$\theta$ (or equivalently $\xi$ and $\eta$) are 
determined by the saddle-point of $F$, we have
\begin{mathletters}%
\begin{eqnarray}
  \epsilon &=& 2 {\rm Im}\,\frac{\partial \tilde S}
  {\partial T}
\\ 
  \nu &=& 2 {\rm Im}\,\frac{\partial \tilde S}
  {\partial \theta} \ ,
\end{eqnarray}
\end{mathletters}%
an alternate expression for the energy and
particle number that can be used as a consistency 
check. 

In this section we have presented the formalism
of Rubakov, Son, and Tinyakov\cite{rst} in some 
detail. The case of a single real scalar field has 
been used for purposes of illustration only, and
the method is easily extended to more complicated
theories. In most instances, and in particular 
for the standard model, the relevant saddle-point 
solutions must be obtained computationally. The 
function $F(\epsilon, \nu)$ can then be determined, 
and the cross section (\ref{sigf}) calculated. 
Finding these complexified saddle-points, however, 
is a formidable numerical challenge, and in the 
next subsection we outline some of our progress 
towards this goal and we present a few initial 
results. 

\subsection{Some Initial Computational Results}

Together with Peter Tinyakov, we are presently 
engaged in the formidable numerical task of 
finding the Euclidean-Minkowski hybrid solutions 
and extracting the baryon number violating 
cross sections. Our approach in the classically
forbidden regime is to map the constant 
\hbox{$F$-contours} by exploring the two 
parameter family of solutions determined by 
$\theta$ and $T$. As this investigation has 
just begun, we shall only present some 
preliminary results and briefly discuss our 
future plans. The numerical approach in the 
classically allowed domain above the
sphaleron barrier is quite different and
involves exploring the $\epsilon$-$\nu$ 
plane using Monte Carlo sampling techniques, 
and we shall present the details of this 
separate investigation in the next section.

As previously stated, we are considering the 
standard model with the Weinberg angle set to 
zero. The resulting spontaneously broken $SU(2)$ 
gauge theory possesses all the relevant physics 
while undergoing notable simplification. The action 
for the bosonic sector of this theory is
\begin{equation}
\label{fourAction}
S = \int dx ^4 ~ \left\{- {1 \over 2} {\rm Tr}\,F_{\mu \nu}
    F^{\mu \nu} + D_{\mu} \Phi^\dagger D^{\mu} \Phi  - \lambda
   (\Phi^\dagger \Phi -1 )^2 \right\} \ ,
\end{equation}
where the indices run from $0$ to $3$ and where
\begin{eqnarray}
  F_{\mu\nu} &=& \partial_\mu A_\nu -  \partial_\nu A_\mu
  - i [A_\mu,A_\nu] 
\\
  D_\mu \Phi &=&  (\partial_\mu - i A_\mu) \Phi \ .
\end{eqnarray}
We use the standard metric $\eta={\rm diag}(1,-1,-1,-1)$,
and have eliminated several constants by a suitable choice 
of units. We have also set $g=1$, but when needed  we 
shall restore the gauge coupling to its physical value of 
$g=0.652$. For our numerical work we take $\lambda=0.1$,
which corresponds to a Higgs mass of about
$M_H= 72 \, {\rm GeV}$. 

To yield a computationally manageable system, we work
in the spherical {\it Ansatz} of Ref.~\cite{ry88} in which 
the gauge and Higgs fields are parameterized in terms 
of six real functions $a_0\, ,\,a_1\, ,
\, \alpha\, , \, \beta\, , \, \mu\, , \, {\rm and}\ 
\nu\ {\rm of}\ r\ {\rm and}\ t$:
\begin{mathletters}%
\label{sphansatz}
\begin{eqnarray}
\label{sphao}
  A_0({\bf x},t) &=& \frac{1}{2 } \, a_0(r,t)
  \lpmb\sigma \cdot {\bf\hat x}
\\
\label{sphai}
  A_i({\bf x},t) &=& \frac{1}{2 } \, \big[a_1(r,t)
  \lpmb\sigma \cdot {\bf\hat x}
  \hat  x^i+\frac{\alpha(r,t)}{r}(\sigma^i- \lpmb\sigma
  \cdot {\bf\hat x}\hat x^i)
  +\frac{1+\beta(r,t)}{r}\epsilon^{ijk}\hat x^j\sigma^k\big]
\\
\label{sphh}
  \Phi({\bf x},t) &=&   [ \mu(r,t) + i \nu(r,t)\lpmb\sigma
  \cdot {\bf\hat x} ] \hat\xi  \ ,
\end{eqnarray}
\end{mathletters}%
where ${\bf \hat x}$ is the unit three-vector in the radial 
direction and $\hat\xi$ is an arbitrary two-component complex 
unit-vector. For the \hbox{4-dimensional} fields to be regular 
at the origin, $a_0$, $\alpha$, $a_1 - \alpha/r$, $(1+\beta)/r$ 
and $\nu$ must vanish like some appropriate power of $r$ as $r 
\to 0$.

These spherical configurations reduce the system
to an effective \hbox{1+1 dimensional} theory on 
the spatial half-line. The action of the reduced
system follows by inserting (\ref{sphansatz}) 
into (\ref{fourAction}), and after some algebra
one obtains\cite{ry88}
\begin{eqnarray}
\label{effAction}
\nonumber
  S =  4\pi \int dt\int^\infty_0dr  &&\bigg[-\frac{1}{4}
  r^2f^{\mu\nu}f_{\mu\nu}+D^\mu \chi^* D_\mu \chi
  + r^2 D^\mu\phi^* D_\mu\phi
\\ 
  && -\frac{1}{2 r^2}\left( ~ |\chi |^2-1\right)^2
  -\frac{1}{2}(|\chi|^2+1)|\phi|^2 -  {\rm Re}(i \chi^* \phi^2)
\\ \nonumber
  && -\lambda  \, r^2 \, \left(|\phi|^2 - 1\right)^2 ~ \bigg] \ ,
\end{eqnarray}
where the indices now run from $0$ to $1$ and 
are raised and lowered with
$\eta_{\mu\nu}={\rm diag}(1,-1)$, and where
\begin{mathletters}%
\begin{eqnarray}
\label{defConva}   
  f_{\mu\nu}&=& \partial_\mu a_\nu-\partial_\nu a_\mu\
\\
\label{defConvb} 
  \chi &=&\alpha+i \beta
\\
\label{defConvc} 
  \phi &=& \mu+i \nu\
\\
\label{defConvd}  
  D_\mu \chi &=& (\partial_\mu- i   \, a_\mu)\chi
\\
\label{defConve}
  D_\mu \phi&=& (\partial_\mu - \frac{i}{2}  \, a_\mu)\phi\ .
\end{eqnarray}
\end{mathletters}%
This is an effective \hbox{2-dimensional}
$U(1)$ gauge theory spontaneously broken
by two scalar fields. It possesses the same rich 
topological structure as the full \hbox{4-dimensional} 
theory and provides an excellent testing ground for 
numerical exploration.

In the rest of this section, we examine 
spherically symmetry Euclidean-Minkowski 
solutions lying on the $ABCD$ time-contour of 
Fig.~1, satisfying the aforementioned boundary 
conditions. In particular, the real gauge field 
$a_\mu$ becomes complex and obeys the boundary
condition
\begin{eqnarray}
  a_\mu(k) &=& 
    g_{\mu ,  k} \,e^{-i\omega_k \tau} 
  + e^\theta \, g^*_{\mu , k} \,
  e^{i\omega_k \tau}    \quad{\rm on\   AB \  as\ }
  \tau={\rm Re} \, t \to-\infty \ ,
\end{eqnarray}
where we have absorbed the factor involving
$\omega_k$ into the definition of the amplitudes. 
For each value of the space-time index $\mu$, 
we are thus searching for two independent 
real degrees of freedom (as the saddle-point
solution $a_\mu$ is complex). The real and the 
imaginary components of the complex fields 
$\chi$ and $\phi$ may be treated in a similar 
manner, giving a total of four real degrees 
of freedom for each field.  We are thus looking 
for two independent complex fields $\chi$ 
and $\bar\chi$, with boundary conditions
\begin{eqnarray}
  \chi(k) &=& 
   f_k \,e^{-i\omega_k \tau} 
  + e^\theta \, h_k \,e^{i\omega_k \tau}   
\\ \nonumber
  \bar\chi(k) &=& 
  h_k^* \,e^{-i\omega_k \tau} 
  + e^\theta \, f_k^* \,e^{i\omega_k \tau}  
  \quad{\rm on\   AB \ as\ }\tau\to-\infty \ ,
\end{eqnarray}
and for two complex fields $\phi$ and $\bar\phi$ 
with similar boundary conditions. When $\theta=0$,
the solution becomes ``real'' along the entire 
$ABCD$ contour, in the sense that $a_\mu$ is real, 
while  $\bar\chi=\chi^*$ and $\bar\phi=\phi^*$. 

In a future publication we plan to numerically 
find these solutions and to explore the behavior 
of the suppression function $F(\epsilon,\nu)$ 
throughout much of the \hbox{$\epsilon$-$\nu$} 
plane. In this paper, however, we restrict 
ourselves to the periodic instantons of 
Ref.~\cite{krt} for which $\theta=0$. 

As path $EF$ of Fig.~2 illustrates, zero
energy instantons 
are Euclidean solutions that interpolate 
between consecutive vacua. In contrast, 
periodic instantons are Euclidean solutions 
that interpolate between configurations of 
nonzero energy lying in adjacent topological 
sectors, as in path $E'F'$ of Fig.~2. They 
have nonzero energy, finite periods, and 
possess  turning points (located at  $E'$ 
and $F'$ in the Figure). Given a periodic 
instanton, we may choose the parameter $T$ 
of Fig.~1 to coincide with the corresponding 
period, and hence we may take $C$ and $B$ 
to be the turning points $E'$ and $F'$ (with 
a time separation of $iT/2$).

\begin{figure}
\centerline{
\epsfxsize=110mm
\epsfbox[0 0 640 427]{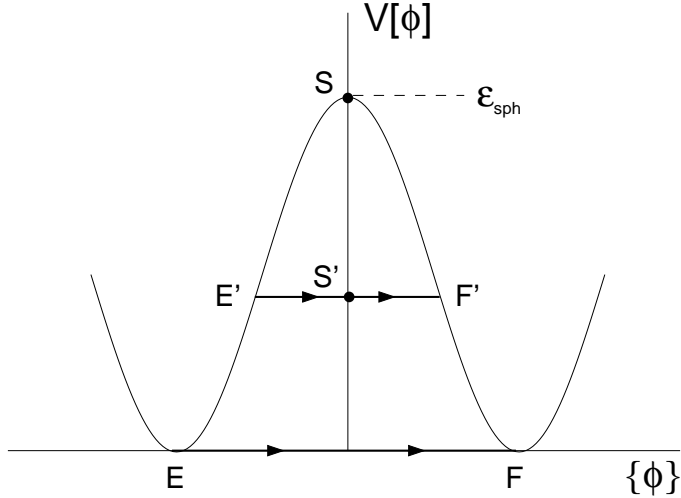}
}
\caption{\tenrm
  The instanton ($EF$), the sphaleron (S), 
  and the periodic instanton ($E'F'$). 
  The horizontal axis represents the infinite 
  dimensional field space, and the vertical 
  axis marks the potential energy of a
  corresponding field configuration. 
}
\end{figure}

We now wish to numerically find the periodic
instanton along the imaginary time axis. These
solutions are real along the Euclidean axis, in 
the sense that $a_\mu$ is real, while 
$\bar\chi=\chi^*$ and $\bar\phi=\phi^*$. 
Moreover, since  the time derivatives vanish 
at $B$ and $C$,  the periodic instanton remains 
real when continued both to the real  axis 
and to the contour  $AB$ (consistent with 
the vanishing of  $\theta$). 

While finding these solutions is less
challenging than obtaining the general 
saddle-points with nonzero $\theta$, 
we must still resort to computational 
methods. However,  before describing 
our numerical approach, it is useful to 
examine two instances in which $F$ 
can be found analytically. Both the
usual zero energy instanton and the 
sphaleron can be viewed as limiting 
cases of periodic instantons, and they 
are illustrated in Fig.~2. The instanton 
represents a vacuum-to-vacuum 
tunneling event, and as such lies at 
the origin of the $\epsilon$-$\nu$ plane, 
thus giving $F(0,0)=-2\,{\rm Im}\,
S_{\rm inst}=-16\pi^2$. 
The sphaleron, on the other hand, has
both nonzero energy and particle
number, $\epsilon_{\rm sph}$ and 
 $\nu_{\rm sph}$ respectively\footnote[2]
{Technically, since the sphaleron is a static 
nonlinear configuration, it does not have 
a particle number; however, a gently perturbed 
sphaleron will decay into a state with a well 
defined particle number  $\nu_{\rm sph}$. It is
the decaying sphaleron, or rather the time 
reversed solution, that we are actually speaking 
of here. To give a feeling for the numbers 
involved, when $g=1$ and $\lambda=0.1$, 
the sphaleron energy is $\epsilon_{\rm sph}=
2.5447$ and the asymptotic particle number 
is $\nu_{\rm sph}=1.7478$ (in physical units
with $g=0.065$, 
$E_{\rm sph}\sim 10\,{\rm TeV}$ and $N_{\rm sph}
\sim 50$).}. It is analytic over the entire 
complex-$t$ plane, with a contribution
to ${\rm Im}\, S$ only along
$BC$ in Fig.~1, and hence  
$F(\epsilon_{\rm sph},\nu_{\rm sph})= 
\epsilon_{\rm sph} T -2\,{\rm Im}\,
S_{\rm sph} =0$. 

We now have the value of $F$ at two key 
points in the $\epsilon$-$\nu$ plane. The 
former gives the usual low energy 't Hooft 
suppression of the baryon number 
violation rates, while the latter yields 
unsuppressed rates at the sphaleron
energy, albeit for very large incident 
particle number (as in a thermal plasma 
in the early universe). All other periodic 
instantons lie along a line connecting 
these two points, and $F$ monotonically 
increases from $-16\pi^2$ to zero as we 
traverse this line from the origin to 
the sphaleron. We now concentrate
on finding these solutions and their
corresponding values of $F$. 

It is convenient to work in the $a_0=0$ 
gauge, and to shift the zero of time so 
that turning points $C$ and $B$ are 
located at $t=-i T/4$ and $t=i T/4$, 
respectively. That is to say,
\begin{mathletters}%
\label{Tconditions}
\begin{eqnarray}
  \dot\chi(r,iT/4) &=& 0 
\\
  \dot\phi(r,iT/4) &=& 0 
\\
  \dot a_1(r,iT/4) &=&  0 \ ,
\end{eqnarray}
\end{mathletters}%
with vanishing time derivatives also at 
$t=-iT/4$.  Since the fields merely retrace 
their steps after the turning points, it 
is sufficient to find the periodic instanton 
only over the half-period from $t=-i T/4$ 
and $t=i T/4$ between consecutive turning
points, which will considerably reduce our 
computational effort. In fact, we can 
be even more economical by exploiting an 
additional symmetry akin to time invariance, 
and it will then suffice to find the periodic 
instanton only over the quarter period, from 
$t=0$ to $t=i T/4$. 

For any solution $\chi(r,t)$, $\phi(r,t)$, 
and $a_1(r,t)$ to the classical equations 
of motion, we may construct another solution 
given by 
\begin{mathletters}%
\label{effprime}
\begin{eqnarray}
  \chi'(r,t) &=& - \chi^*(r,-t)
\\
  \phi'(r,t) &=& - \phi^*(r,-t)
\\
  a'_1(r,t) &=& - a_1(r,-t) \ .
\end{eqnarray}
\end{mathletters}%
This can be traced to a combination of 
parity symmetry (in the full \hbox{4-dimensional} 
theory), $\phi \to -\phi$ invariance, and 
simple time reflection $t \to -t$. In terms 
of the reduced theory, this is none other 
than the \hbox{2-dimensional} time reversed 
solution (which should not be confused 
with \hbox{4-dimensional} time reversal).

A corresponding time reversed solution 
may be obtained from any given solution
by appropriately changing the initial
conditions in a manner dictated by 
(\ref{effprime}). However, if the
initial conditions take the form,
\begin{mathletters}%
\label{icforminus}
\begin{eqnarray}
  \chi(r,0) &=& {\rm imaginary} \hskip1cm 
  \dot\chi(r,0) = {\rm real} 
\\
  \phi(r,0) &=& {\rm imaginary} \hskip1cm 
  \dot\phi(r,0) = {\rm real} 
\\
  a_1(r,0) &=& 0  \ ,
\end{eqnarray}
\end{mathletters}%
(with no restriction on the time derivative of
$a_1$), then the solution will be invariant under
time reversal: \hbox{$\chi'=\chi$}, 
\hbox{$\phi'=\phi$}, and \hbox{$a'_1=a_1$}, 
i.e.
\begin{mathletters}%
\label{minussym}
\begin{eqnarray}
  \chi(r,-t) &=& - \chi^*(r,t)
\\
  \phi(r,-t) &=&  -\phi^*(r,t)  
\\
  a_1(r,-t) &=&  -a_1(r,t) \ ,
\end{eqnarray}
\end{mathletters}%
and in this case, we only need to look for 
solutions over the quarter period from $t=0$ 
to $t=i T/4$, satisfying the initial and 
final conditions (\ref{Tconditions}) and 
(\ref{icforminus}).

We must also choose boundary 
conditions at the origin and spatial 
infinity in such a way as to ensure 
regularity of the corresponding 
\hbox{4-dimensional} fields. For the
$\chi$ field, we must take $\chi(0,t)=-i$,
and we may choose a gauge in which
$\chi(r,t)=i$ as $r\to\infty$. Thus, as 
the initial  $\chi$-configuration is 
pure imaginary, it will always have 
a zero for some nonzero value of $r$.
As will be discussed
more thoroughly in the next
section, the sphaleron is a spherical
configuration with vanishing $a_\mu$
and with pure imaginary $\chi$ and 
$\phi$; furthermore, like the initial 
configuration, its $\chi$-field possesses 
a zero at some radius $r$ away from the 
origin. Hence, the  initial 
configurations (\ref{icforminus}) 
will not lie far from the sphaleron, 
as indicated by the close  proximity 
of $S'$ and $S$ in Fig.~2.

To find the periodic instanton
numerically, we  place the system 
in a box of spatial 
extent $L_r$ and time extent $L_t$, 
and then discretize the action 
(\ref{effAction}) using the standard 
techniques of lattice gauge theory.
The space-time grid has lattice sites 
at $(i \Delta r,j \Delta t)$, 
where $i=0 \cdots N_r$ and $j=0 
\cdots N_t$. The fields $\chi$ and 
$\phi$ become discrete variables 
defined on these sites, while $a_1$ 
is defined on the space-links and 
time-sites (in a gauge where $a_0$ 
does not vanish, it  is defined on  
the space-sites and time-links). We 
take $N_r=64$, $N_t=40$, with $dr=0.05$,
and impose (\ref{Tconditions}) and
(\ref{icforminus}) on the lower and 
upper time slices $j=0$ and $j=N_t$ 
respectively. The parameter $\Delta t$, 
which is taken between $0.02$ 
to $0.04$, controls the period of the 
periodic instanton through 
$T= 4 \Delta t N_t$.

Starting from an initial guesses along the 
Euclidean axis satisfying the appropriate 
boundary conditions, we search for a 
minimum of the action using the method 
of conjugate gradients. Like the naive 
gradient descent, the conjugate gradient 
algorithm chooses its descent direction
based upon the gradient. A new guess is
then selected lying further down the 
slope, and the algorithm is repeated. 
With each new iteration, the configuration
finds itself closer and closer to the
local minimum, and eventually one can
approximate the extremum to within the
desired tolerance. 
The advantage of the conjugate gradient 
method over a simple gradient descent, 
is that the former achieves a much more 
rapid convergence rate by judiciously 
shifting the direction of descent slightly 
away from the gradient. As this method 
is rather standard,  we shall not describe 
it in any more detail. 

A straightforward application of the algorithm, 
however, yields little success. Periodic 
instanton may be found in this manner,  but 
unless one starts extremely close to a solution,
one typically relaxes to the static sphaleron. 
The two turning points of the initial guess, 
$E'$ and $F'$ in Fig.~2, slide onto
the sphaleron-like configuration $S'$, which 
in turn eventually relaxes to the sphaleron 
itself. To avoid this, we add an additional
term to the action that tends to pin the turning
points, thereby halting the collapse into the
sphaleron. We do this by minimizing an
effective action of the form
\begin{eqnarray}
\label{seffvnt}
  S_{\rm eff} = S + w_t \, (V_{N_t}-v_0)^2 \ ,
\end{eqnarray}
where $V_{N_t}$ is the potential energy
on the final time slice, $w_t$ is a 
weighting factor, and $v_0$ is called 
the turning-point energy parameter. While
the second term in (\ref{seffvnt}) renders 
collapse to the sphaleron energetically 
unfavorable, in general the minima 
of $S_{\rm eff}$ are not the solutions 
we seek. However, if we choose the
parameter $v_0$ such that $V_{N_t}=v_0$, 
the minima of $S_{\rm eff}$ and $S$
coincide. Thus, by adjusting  $v_0$ 
accordingly, we can find periodic 
instantons that do not collapse to 
the sphaleron. 

The effective action  (\ref{seffvnt}) 
still typically fails to yield nontrivial 
periodic instantons. While the
second term in (\ref{seffvnt}) pushes
the gradient search away from the
static sphaleron, there is still
another unstable direction. Rather 
than converging to periodic instanton
solutions, most initial guesses
shrink to zero size. This is related 
to the fact that the standard model, 
with a nonzero vacuum expectation value, 
breaks conformal invariance and strictly
speaking does not support instanton 
solutions (i.e., they shrink to zero 
size). We can remedy the situation by 
adding another term to the action that 
pins the zero of $\chi(r,0)$, thereby
halting the collapse of the configuration.
We now consider the effective action
\begin{eqnarray}
\label{seffzero}
  S_{\rm eff} = S + w_t \, (V_{N_t}-v_0)^2 +
  w_{\rm zero} \left[ (1-\alpha) \chi_{i,0}+ 
\alpha \chi_{i+1,0}\right]^2 \ ,
\end{eqnarray}
where $w_{\rm zero}$ is another weighting
factor. Just as before, we seek to minimize
$S_{\rm eff} $, but in addition to adjusting
$v_0$ so the second term of (\ref{seffzero}) 
vanishes, we also vary $\alpha$ and $i$ 
to give a vanishing third term (and hence the
zero of $\chi(r,0)$ occurs at $r_0=(i+\alpha)
\Delta r$). 

A conjugate gradient minimization 
of the effective action (\ref{seffzero}), coupled 
with the two parameter search over $v_0$ and 
$\alpha$, is a very effective method for obtaining
periodic instantons. In this paper we only have 
space to present a typical solution, shown in Fig.~3, 
and a comprehensive treatment of periodic instantons 
in the $\epsilon$-$\nu$ plane must wait for a future 
publication.

\vskip1cm
\begin{figure}
\centerline{
\epsfxsize=70mm
\epsfbox{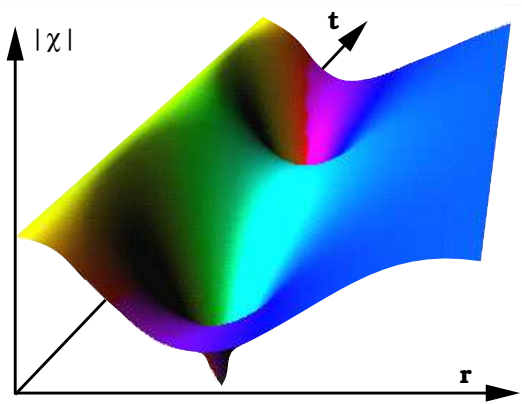}
}
\vskip0.25cm
\caption{\tenrm
Periodic Instanton: a full period of the $\chi$-field. 
The modulus of the field is represented by the height 
of the surface, while shades of gray code the phase. 
}
\end{figure}

\section{The Classically Allowed Domain}
\hspace{1cm}
We now examine the complimentary
classically allowed regime above the 
sphaleron barrier, in which the solutions 
are purely real and propagate in
Minkowski space-time. Finding these
solutions is less computationally 
demanding than solving the tunneling 
problem of the previous section,
while still yielding considerable information 
about baryon number violation. 
Spatial limitations prevent us from giving 
a full blown treatment of our numerical 
investigation, and the reader is referred 
to Ref.~\cite{rs} for complete details. 
But the basic idea is that if a topology 
changing classical solution with small incident 
particle number could be found, this would be 
a strong indication that baryon number violation 
would be observable in high energy two-particle 
collisions. Conversely, if there are no 
small-multiplicity topology changing solutions, 
then it is unlikely that the rates become 
exponentially unsuppressed.  

This can be made more precise in the following 
manner. Because of energy dissipation, the 
system will asymptotically approach vacuum 
values and will consequently linearize in 
the past and future. Field evolution then
becomes a superposition of normal mode 
oscillators with amplitudes $a_n$, which 
allows us to define the asymptotic particle 
number $\nu = \sum |a_n|^2$. 
Furthermore, since the fields approach 
vacuum values in the infinite past and 
future, the winding numbers of the asymptotic 
field configurations are also well defined, 
and fermion number violation is given by 
the change in topology of these vacua\cite{fggrs}. 
Because of the sphaleron barrier, classical
solutions that change topology must have
energy $\epsilon$ greater than that of the
sphaleron. The problem we would like to solve, 
then, is whether the incident particle number 
$\nu$ of these solutions can be made arbitrarily
small. That is to say, we wish to map the
region of topology changing classical solutions
in the $\epsilon$-$\nu$ plane. 

We could easily parameterize incoming
configurations in terms of small perturbations 
about a given vacuum, but it would be 
extremely difficult to choose the parameters 
to ensure a subsequent change in winding
number. This is because topology changing 
classical solutions must pass over the sphaleron 
barrier at some point in their evolution, which 
is extremely difficult to arrange by an 
appropriate choice of initial conditions. 
So computationally, 
we pursue a different strategy. We will evolve
a configuration near the top of the sphaleron
barrier until it linearizes and the particle number
can be extracted. The time reversed solution,
then, has a known incident particle number and 
will typically pass over the sphaleron barrier
thereby changing topology. 
Of course we have no obvious control over the
asymptotic particle number of the initial sphaleron-like 
configuration; however, by using suitable stochastic
sampling techniques, we can systematically
lower the particle number while ensuring a change
of topology. This will allow us to explore the
$\epsilon$-$\nu$ plane and map the region of topology
change, the lower boundary of which should tell 
us a great deal about baryon number violation 
in high energy collisions. 

\subsection{Topological Transitions}

Let us now put some flesh on the bones of the
above discussion. As in the previous section,
we still consider the standard model with zero 
Weinberg-angle, defined by action (\ref{fourAction}). 
As before, the coupling constant has been set 
to unity, but when needed it will be restored 
to its physical value of $g=0.652$. We also take 
$\lambda=0.1$, which corresponds to a Higgs mass 
of about $M_H= 72 \, {\rm GeV}$. Again we restrict 
ourselves to the spherical {\it Ansatz} 
(\ref{sphansatz}), and examine the effective
1+1 dimensional $U(1)$ theory (\ref{effAction}). 

Before investigating classical solutions of this
effective theory, it is useful to first explore  
its topological structure, which is very 
similar to that of the full \hbox{4-dimensional} 
theory. Vacuum states of the effective 
\hbox{2-dimensional} theory are characterized by 
$|\chi |= |\phi|=1$ and $i\chi^* \, \phi = -1$ 
(as well as $D_\mu \chi= D_\mu \phi=0)$. The 
vacua then take the form
\begin{eqnarray}
\nonumber
 a_{\mu \, \rm vac} &=&  \partial_\mu \Omega
\\ \label{chiphiavac}
  \chi_{\rm vac} \, &=& -i \, e^{i \Omega}
\\ \nonumber
  \phi_{\rm vac} \, &=& \pm \, e^{i \Omega/2} \ ,
\end{eqnarray}
where  the gauge function $\Omega=\Omega(r,t)$ 
is required to vanish at $r=0$ to ensure regularity 
of the \hbox{4-dimensional}  fields. Furthermore, 
like the full  \hbox{4-dimensional} theory, these
vacua still possess nontrivial topological 
structure. Compactification of \hbox{3-space} requires
that $\Omega(r,t) \to 2\pi n$ as $r \to \infty$, in 
which case the winding number of such vacua 
in the $a_0=0$ gauge is simply the integer 
$n$.  Note that as $r$ varies from zero to infinity, 
$\chi$ winds $n$ times around the unit  circle 
while $\phi$ only winds by half that amount. 

Since the winding number is a topological 
invariant, a continuous path connecting two 
inequivalent vacua must at some point 
leave the manifold of vacuum configurations. 
Along this path there will be a configuration 
of maximal energy, and of all such maximal
energy configurations there exists a unique
one of minimal energy\cite{KM}. This 
configuration is called the sphaleron and 
may conveniently be parameterized by
\begin{eqnarray}
\label{sphSphal}
\nonumber 
  a^\mu_{\rm sph}(r)&=&0
\\ 
  \chi_{\rm sph}(r) &=&i [2 f(r)-1] 
\\ \nonumber  
  \phi_{\rm sph}(r)&=&i h(r) \ ,
\end{eqnarray}
where the profile functions $f$ and $h$ vanish 
at $r=0$, and tend to unity as  $r\to \infty$
and are otherwise determined by minimizing
the energy functional. The sphaleron energy 
is approximately $M_w/g^2 \sim 10~{\rm TeV}$, 
or $\epsilon=4\pi \, (2.54)$ for $\lambda=0.1$
in the units we are using (and it depends 
very weakly on the Higgs mass).

\begin{figure}
\centerline{
\epsfxsize=100mm
\epsfbox{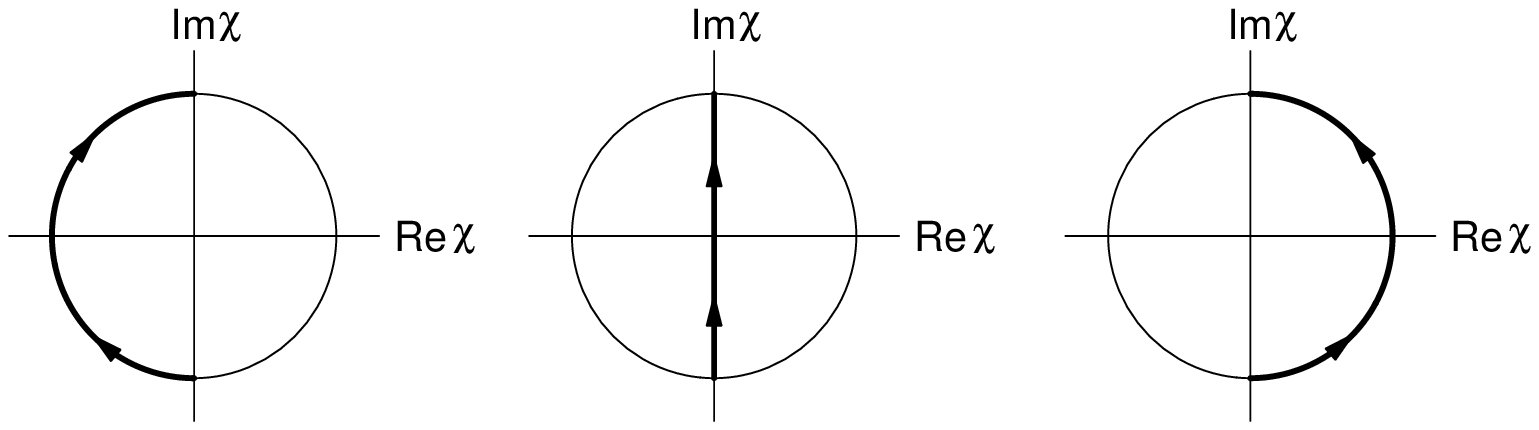}
}
\centerline{
\epsfxsize=105mm
\epsfbox{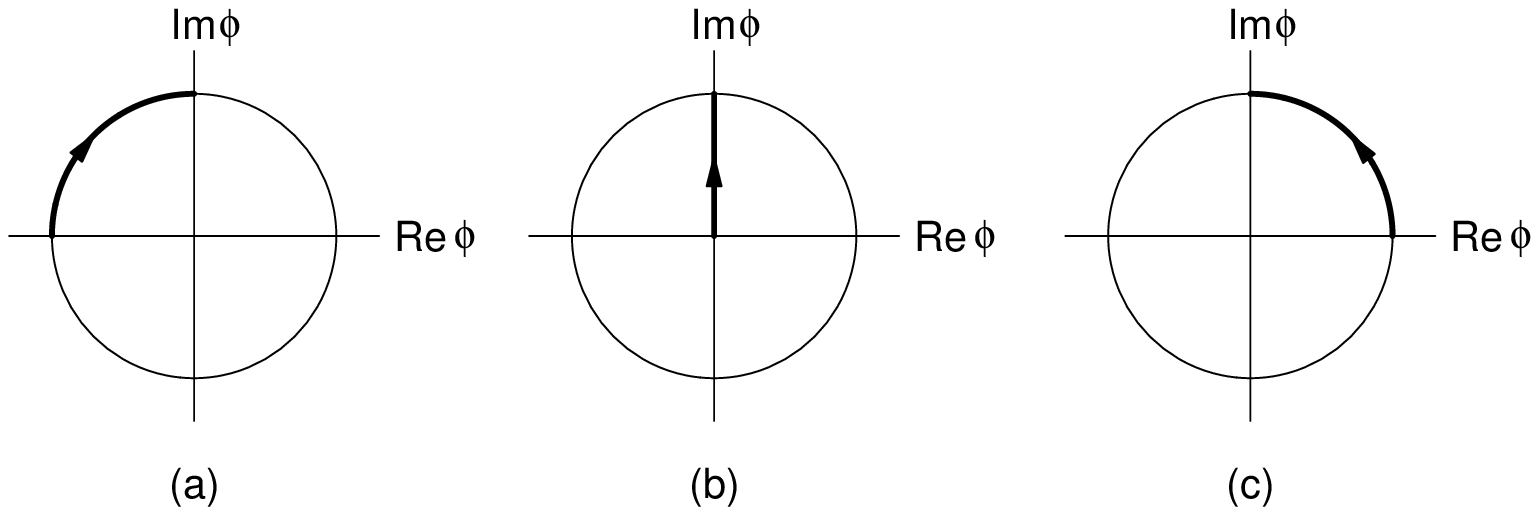}
}
\caption{\tenrm
The $\chi$ and $\phi$ fields for a vacuum-to-vacuum 
topology changing transition in a gauge inconsistent
with compactified \hbox{3-space}.
The scalar fields are traced in the complex plane as 
the spatial coordinate spans the entire axis.  
Figs.~(a) and (c) represent two inequivalent topological
vacua while (b) is the sphaleron barrier separating them.
}
\end{figure}

While the form of the sphaleron given by (\ref{sphSphal}),
in which $a_\mu$ vanishes and $\chi$ and $\phi$ 
are pure imaginary, is convenient for numerical 
work, it does have a 
slight peculiarity. Recall that compactification 
of \hbox{3-space} requires the gauge function $U$ 
to approach an even multiple of $2\pi $ as 
\hbox{$r \to \infty$}. It is possible to relax 
this restriction, and it will often be convenient 
to choose a gauge with $U \to (2n+1) \pi$ as 
\hbox{$r \to \infty$}, in which case 
\hbox{$\chi_{\rm vac} \to i$} and \hbox{$\phi_{\rm vac} 
\to \pm i$}. This is precisely the large-$r$ boundary 
condition of the sphaleron, which illustrates 
that (\ref{sphSphal}) is inconsistent with 
spatial compactification. There is of course 
nothing wrong with this, and a topological transition 
of unit winding number change in this gauge
is illustrated in Fig.~4. Rather than $\chi$ winding
once around the unit circle, it instead winds 
over the left hemisphere before the transition
and over the right after the transition.  The total 
phase change is still $2\pi$, as it must be since 
this is a gauge invariant quantity. 

Throughout most of this paper we shall use a gauge 
consistent with (\ref{sphSphal}) in which space
cannot be compactified. From a computational
perspective, this will allow perturbations about 
the sphaleron to be more easily parameterized. 
There will, however, be times when it is more 
convenient to impose spatial compactification, 
but we will always alert the reader to such 
a change of gauge.

\subsection{Classical Evolution} 

So far we have primarily been considering topology
changing sequences of configurations, not 
necessarily solutions of the equations of motion. 
We now turn to the classical evolution of the
system. We will consider solutions that 
linearize in the distant past and future, and  
hence those that asymptote to specific topological 
vacua. This allows us to define the incident particle
number, and it makes clear what is meant by 
the topology change of a classical solution 
(namely, the change in winding number of 
the asymptotic vacua). 

The field equations are coupled nonlinear
particle differential equations and must
be solved computationally on the lattice. But 
before we present our  numerical procedure, 
we first formulate the problem in the continuum. 
The equations of motion resulting from the 
action (\ref{effAction}) are
\begin{mathletters}%
\label{fieldEqs}
\begin{equation}
\label{fEq}
  \partial^\mu(r^2f_{\mu\nu})=i \left[D_\nu \chi^*\chi-\chi^*
  D_\nu\chi \right] + \frac{i }{2}\,   \, r^2 \left[D_\nu
  \phi^*\phi-\phi^*D_\nu\phi\right]
\end{equation}
\begin{equation}
\label{chiEq}
  \left[D^2+\frac{1}{r^2}(|\chi|^2-1) + \frac{1}{2}\,  
  |\phi |^2 ~ \right]\chi=-\frac{i}{2}\,  \, \phi^2
\end{equation}
\begin{equation}
\label{phiEq}
  \left[D^\mu r^2 D_\mu+\frac{1}{2}(|\chi|^2+1) +
  2\lambda r^2 \left(|\phi|^2- 1 \right)
  \right] \phi= i \, \chi \phi^* \ .
\end{equation}
\end{mathletters}%
The $\nu=0$ equation in (\ref{fEq}) is not dynamical 
but is simply the Gauss's law constraint. 

To solve
these equations, we must supplement them with 
boundary conditions. The conditions at $r=0$ can 
be derived by requiring the \hbox{4-dimensional} 
fields to be regular at the origin. The behavior as 
$r\to 0$ must be
\begin{mathletters}%
\label{zerobc}
\begin{eqnarray}
  a_0 &=& a_{0,1} r +  a_{0,3} r^3 +\dots 
\\ 
  a_1&=& a_{1,0} + a_{1,2} r^2 + \dots 
\\
  \alpha&=& \alpha_1 r + \alpha_3 r^3  +\dots 
\\
  \beta&=&-1 + \beta_2 r^2 + \dots 
\\
  \mu&=&\mu_0 + \mu_2 r^2 + \dots 
\\ 
  \nu&=&\nu_1 r + \nu_3 r^3 + \dots \ ,
\end{eqnarray}
\end{mathletters}%
where the coefficients of the $r$-expansion are 
undetermined functions of time. The $r$-behavior
of the various terms are determined by the
requirement that it has the appropriate power
of \hbox{$r=(x^2+y^2+z^2)^{1/2}$} to render
the \hbox{4-dimensional} fields analytic in terms 
of $x$, $y$ and $z$. For example, $a_0$ must
be odd in $r$ since $A_0$ is proportional to 
$a_0 \lpmb{\sigma} \cdot {\bf \hat x}= (a_0/r) 
\lpmb{\sigma} \cdot {\bf x}$. In terms of $\chi$
and $\phi$ the boundary conditions at $r=0$ 
become 
\begin{mathletters}%
\label{newzerobc}
\begin{eqnarray}
  a_0(0,t) &=& 0 
\\ 
  \chi(0,t)&=&- i  
\\
\label{rephibc}
  {\rm Re}\, \partial_r\phi(0,t)&=&0  
\\
\label{imphibc} 
  {\rm Im} \, \phi(0,t)&=& 0 \ .
\end{eqnarray}
\end{mathletters}%
There is another $r=0$ boundary condition which 
arises from the requirement that $a_1-\alpha/r$ 
be regular as $r \to 0$. In the notation of
(\ref{zerobc}), this condition can be 
written as $a_{1,0}=\alpha_1$, and once imposed 
on initial configurations it remains satisfied 
at subsequent times because of Gauss's law. 

We turn now to the large-$r$ boundary conditions.
Finite energy configurations must approach pure
vacuum at spatial infinity, and we may choose a
gauge in which 
\begin{mathletters}%
\label{largerbc}
\begin{eqnarray}
\label{alarger}
  a_\mu(r,t) & \to & 0 
\\
  \chi(r,t)  &\to & i  
\\
  \phi(r,t) &\to& i  
\end{eqnarray}
\end{mathletters}%
as $r\to\infty$. This choice of gauge does not admit
spatial compactification, but nonetheless it is 
numerically convenient since it is consistent 
with the simple parameterization of the sphaleron 
(\ref{sphSphal}).  At times we will choose a gauge 
consistent with spatial compactification in which  
\hbox{$\chi(r,t) \to -i$} and \hbox{$\phi(r,t) 
\to 1$} as $r \to \infty$, but unless otherwise 
specified we will take the large-$r$ boundary 
conditions to be (\ref{largerbc}).

The field equations (\ref{fieldEqs}), together
with boundary conditions (\ref{newzerobc}) and 
(\ref{largerbc}), may now be used to evolve 
initial profiles and to investigate their subsequent
topology change. The evolution is performed by 
discretizing the system using the methods of  
lattice gauge theory, in which we subdivide the 
$r$-axis into $N$ equal intervals of length 
$\Delta r$ with finite extent $L=N \Delta r$ 
(in our numerical simulations we take $N=2239$ 
and $\Delta r = 0.04$). The field theoretic 
system then becomes finite and may be solved 
using standard numerical techniques. 

The  fields $\chi(r,t)$ and $\phi(r,t)$ become 
discrete variables $\chi_i(t)$ and $\phi_i(t)$ associated 
with the lattice  sites $r_i= i \Delta r$ where $i=0 \cdots N$.
The continuum boundary conditions render the variables
at the spatial end-points nondynamical, taking the values
\hbox{$\chi_0=-i$},  \hbox{$\chi_N=i$} and \hbox{$\phi_N=i$}
(the value of $\phi_0$ will be discussed momentarily). The 
time component of the gauge field $a_0(r,t)$ is also 
associated  with the lattice sites and is represented 
by the variables $a_{0,i}(t)$ with $i=0 \cdots N$. We
will usually work in the temporal gauge in which $a_{0,i}=0$, 
and we will not concern ourselves with this degree of freedom. 

The spatial components of the gauge field $a_1(r,t)$
become discrete variables associated with the oriented 
links of the lattice, and we represent them by $a_{1,i}(t) 
\equiv a_i(t)$ located at positions $r_{i+1/2}=(i+1/2)\Delta r$
with $i=0\cdots N-1$. The covariant spatial derivatives
become covariant finite difference operators that are
also associated with the links, e.g.
\begin{equation}
\label{Drdisc}
  D_r \phi \to { \exp[- i a_i\, \Delta r/2] \,
  \phi_{i+1}- \phi_i \over \ \Delta r}
  {}~~~~~~~ i=0 \cdots N-1 \ ,
\end{equation}
where $a_i$ is short-hand notation for  $a_{1,i}$.

It is now straightforward to discretize the action (\ref{effAction})
in a manner that still possesses an exact local gauge invariance. 
But first, we need to state the restriction on $\phi_0(t)$
corresponding to the boundary conditions 
(\ref{rephibc}) and (\ref{imphibc}). Since $a_1$ is real,
we can write these boundary conditions in a 
covariant fashion by requiring the real part of $D_r\phi$
and the imaginary part of $\phi$ to vanish at $r=0$. Using
the discretized operator (\ref{Drdisc}), we can then solve 
this boundary condition for $\phi_0$ to obtain 
\begin{equation}
\label{phi0bc}
  \phi_0 =  {\rm Re\,} [ \exp(-i \,a_0\, \Delta r/2) \phi_1] \ ,
\end{equation}
where $a_0$ is the value of $a_{1,i}$ at $i=0$ and should
not be confused with the time-like vector field.  This now 
allows us to eliminate $\phi_0$ from the list of
dynamical variables.

Finally, the discretized Lagrangian becomes
\begin{eqnarray}
\label{discL}
\nonumber
  L &=&4\pi \, \sum_{i=0}^{N-1} \bigg\{\frac{r^2_{i+1/2} }{2} \,
  \bigg(\partial_0 a_i   -\frac{a_{0,i+1}-a_{0,i}}{\Delta r}
  \bigg)^2 - \frac{ |\exp(-i\,a_i\, \Delta r)
  \chi_{i+1} -\chi_i|^2}{\Delta r^2} \bigg\} \Delta r
\\ \nonumber && \hskip-0.5in
  + \,4\pi \, \sum_{i=1}^{N-1} \bigg\{ |(\partial_0-i
  a_{0,i})\chi_i|^2+r^2_i \, |(\partial_0-{i a_{0,i} 
  \over 2})\phi_i|^2 - r^2_{i+1/2}  \, \frac{|\exp({-i\,a_i\,
  \Delta r / 2}) \phi_{i+1} -\phi_i|^2}{\Delta r^2 }
\\ &&\hskip-0.5in
  -\frac{1}{2} \, (|\chi_i|^2+1)|\phi_i|^2 -  {\rm Re}(i\chi_i^*
  \phi_i^2) - {1 \over 2 r^2_i}(|\chi_i|^2-1)^2  - \lambda
  r^2_i \, (|\phi_i|^2-1)^2  \bigg\} \Delta r
\\ \nonumber &&\hskip-0.5in
  - \,4\pi \, r^2_{1/2}{[{\rm Im}(\exp(-i\,a_0\, \Delta r / 2)
  \phi_1)]^2 \over  \Delta r} \ ,
\end{eqnarray}
and the system may now be evolved using  standard 
numerical techniques of ordinary differential equations. 
The Lagrangian (\ref{discL}) is actually of a Hamiltonian 
type with no dissipative terms, so it is convenient to 
use the  leapfrog algorithm to perform the numerical 
integration. 

We do not have space to outline this well known 
computational procedure, so instead we simply state 
some of its more attractive features. First, the 
algorithm is second order accurate (i.e.~the error 
{}from time discretization is of order $(\Delta t)^3$ 
in the individual steps and of order $(\Delta t)^2$ 
in an evolution of fixed length $L=N\Delta r$). 
Second, energy is exactly conserved in the linear 
regime, a desirable feature when  pulling out the 
particle number. And finally, the algorithm possesses 
an exact discretized-time invariance, which is important 
since we are interested in obtaining the time reversed 
solutions starting from perturbations about the sphaleron. 
Of course these last two properties hold exactly only 
up to round-off errors, which can be made quite small 
by using double precision arithmetic.

\subsection{The Initial Configuration: Perturbation About
the Sphaleron }

We are now ready to continue our investigation 
into the connection between the incident particle
number of a classical solution and subsequent
topology change. We could proceed by 
parameterizing linear incoming configurations
of known particle number, but it would be extremely
difficult to arrange the classical trajectory to 
traverse the sphaleron barrier. If we failed to 
see topology change for a given initial 
configuration, we could never be sure whether 
it was simply forbidden in principle by the choice 
of incident particle number, or simply because 
the initial trajectory was pointed towards the 
wrong direction in field space. 

To alleviate this difficulty, we have chosen 
to evolve initial configurations at or near the 
moment of topology change, and when the 
linear regime is reached the particle number 
will be extracted in the manner explained 
shortly. The physical process of 
interest is then the time reversed solution 
that starts in the linear regime with known 
particle number and subsequently proceeds 
over the sphaleron barrier. Of course we must
explicitly check whether topology change
in fact occurs, but we have found that 
it usually does. 
Fig.~5 illustrates the numerical evolution 
of the $\chi$ field for a typical topology 
changing solution obtained in this manner. 
The modulus of  $\chi$  is represented
by the height of the surface, while the phase
is color coded (but unfortunately we can only 
reproduce the figure in gray scale).  
We have reverted to a gauge where $\chi_N=-i$ 
and $\phi_N=1$, consistent with spatial compactification,
and in which the incoming state has no winding and
the outgoing state has unit winding number. The
topology change 
is represented by the persistent strip of $2\pi$ 
phase change near the origin after the transition. 
\begin{figure}
\centerline{
\epsfxsize=70mm
\epsfbox{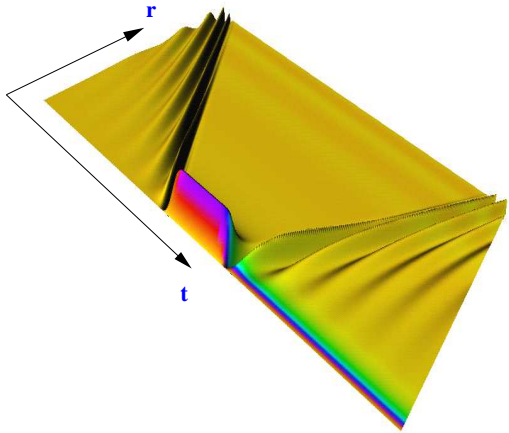}
}
\caption{\tenrm
Topology changing transition: behavior of the 
$\chi$ field obtained the time reversal procedure 
described in the text. The various shades of gray 
code the phase of the complex field. The field 
starts as an excitation about the trivial vacuum, 
passes over the sphaleron and then emerges as an
excitation about the vacuum of unit winding. Note 
the persistent strip of $2\pi$ phase change near 
$r=0$ after the wave bounces off the origin. 
}
\end{figure}

We turn now to parameterizing initial
configurations. For classical solutions  that 
dissipate in the past and future, topology 
change (and hence baryon number violation) 
is characterized by zeros of the Higgs 
field\cite{fggrs}.
For such topology changing solutions in the
spherical {\it Ansatz}, the $\chi$ field, which 
parameterizes the transverse gauge 
degrees of freedom, must also vanish 
at some point in its evolution. 
However, unless the transition
proceeds directly through the sphaleron, 
the zeros of $\phi$ and $\chi$ need not
occur simultaneously, and for convenience 
we shall choose to parameterize the initial 
configuration at the time when $\chi$ 
vanishes for some nonzero $r$. Furthermore,
we can exhaust the remaining gauge 
freedom by taking the initial $\chi$ to be 
pure imaginary. We thus parameterize 
the initial conditions as an expansion in 
terms of some appropriate complete set
with coefficients $c_n$, consistent only with 
the boundary conditions and the requirement
that $\chi$ be pure imaginary with a zero
at some $r>0$. 

We choose to parameterize initial conditions in 
terms of perturbations about the sphaleron given 
by linear combinations of spherical Bessel functions 
consistent with the small-$r$ behavior (\ref{zerobc}). 
We only need the first three functions
\begin{mathletters}%
\begin{eqnarray}
  j_0(x) &=& \frac{\sin x}{x}   
\\
  j_1(x) &=& \frac{\sin x}{x^2} - \frac{\cos x}{x}   
\\
  j_2(x) &=&   \left(\frac{3}{x^3} - \frac{1}{x}\right) \sin x -
  \frac{3}{x^2} \cos x  \ ,
\end{eqnarray}
\end{mathletters}%
since $j_0(x)\sim 1$, $j_1(x)\sim x$ and $j_2(x)\sim x^2$
at small $x$. We also require the perturbations to vanish
at $r=L$ consistent with the large-$r$ boundary conditions
(\ref{largerbc}). We then parameterize
perturbations about the sphaleron in terms of 
$j_{nm}(r)=j_n(\alpha_{nm}r)$ with $n=0,1,2$, 
where $\alpha_{nm}$ with $m=1,2,\cdots\,$ are the 
zeros of $j_n(x)$. We are thus led to parameterize the
initial conditions as

\
\begin{mathletters}%
\label{initconfig}
\begin{eqnarray}
\label{chiparam} 
  \chi(r,0) &=& 
  \chi_{\rm sph}(r) + i \sum_{m=1}^{N_{\rm sph} } \, c_{1 m}\
  j_{2 m}(r) 
\\
\label{phiparam} 
  \phi(r,0) &=& \phi_{\rm sph}(r) + \sum_{m=1}^{N_{\rm sph} } 
  \, c_{2 m}\ j_{0 m}(r)+ i \sum_{m=1}^{N_{\rm sph} } \, 
  c_{3 m}\  j_{1 m}(r)
\\
\label{pichiparam} 
  \dot \chi(r,0) &=& \sum_{m=1}^{N_{\rm sph} } \, c_{4 m}\
   j_{1 m}(r)+ i \sum_{m=1}^{N_{\rm sph} } \, c_{5 m}\  
  j_{2 m}(r)
\\
\label{piphiparam} 
  \dot \phi(r,0) &=&  \sum_{m=1}^{N_{\rm sph} }\, 
  c_{6 m}\ j_{0 m}(r)+ i \sum_{m=1}^{N_{\rm sph} } \, 
  c_{7 m}\  j_{1 m}(r)   
\\
\label{aparam}
  a_1(r,0) &=& \sum_{m=1}^{N_{\rm sph} } \,c_{8 m}\ 
  j_{2 m}(r) \ ,
\end{eqnarray}
\end{mathletters}%
where $\chi_{\rm sph}$ and $\phi_{\rm sph}$ are
the sphaleron profiles, and where the sum is cut
off at \hbox{$N_{\rm sph} \le N$}. To avoid exciting short
wave length modes corresponding to lattice
artifacts, we shall take $N_{\rm sph} \sim N/50$ 
(in our numerical work, $N_{\rm sph} =50$ for 
$N=2239$). 

We have used continuum notation, but (\ref{initconfig})
is to be thought 
of as defining $\chi$ and $\phi$ on the lattice sites 
$r_i$ and $a_1$ on the links $r_{i+1/2}$. The time 
derivative of $a_1$ is to be determined by Gauss's 
law. 

\subsection{Normal Modes and Particle Number}

We are now in a position to discuss the
manner in which the asymptotic particle
number is to be extracted.
Recall that once the system has reached 
the linear regime it can be represented
as a superposition of normal modes,
and the particle number can be defined 
as the sum of the squares of the 
normal mode amplitudes.
Since we have put the system on a
lattice, we should properly calculate
these amplitudes using the exact 
normal modes of the discrete system. 
However, since our lattice is very
dense ($N=2239$ with $\Delta r=0.04$),
it suffices to project onto the normal 
modes of the corresponding continuum 
system of finite extent $L=N\Delta r$,
the advantage being that we can
solve for the continuum normal modes 
analytically. 
We have checked that this procedure 
agrees extremely well with projecting 
onto normal modes of the discrete system 
(obtained numerically), so for clarity 
we present only the continuum modes. 

It is convenient to work in terms of the gauge 
invariant variables of Ref.~\cite{gi}. We write 
the fields $\chi$ and $\phi$ in polar form,
\begin{eqnarray}
  \chi &=& -i \left[ 1+y \right]\, e^{i\theta} 
\\ \nonumber \\
  \phi &=& \left[ 1+{h \over r} \, \right] \, e^{i\eta} \ ,
\end{eqnarray}
where the variables $y$ and $h$ are gauge invariant.
We can also define the gauge invariant angle
\begin{eqnarray}
\xi=\theta-2\eta \ ,
\end{eqnarray}
and in \hbox{1+1} dimensions we can define
a gauge invariant quantity $\psi$ through
\begin{eqnarray}
  r^2 f_{\mu\nu} = - 2 \epsilon_{\mu\nu} \psi \ ,
\end{eqnarray}
where $\epsilon_{01}=+1$ and $\mu$, $\nu$ run 
over $0$ and $1$. Rather than working with the 
six gauge-variant degrees of freedom $\chi$, 
$\phi$ and $a_\mu$ we use the four gauge 
invariant variables $h$, $y$, $\psi$ 
and $\xi$. 

We wish to find the equations of motion for 
small linearized fluctuations about the vacuum. 
In gauge invariant coordinates the vacuum takes 
the form $h_{\rm vac}=y_{\rm vac}=\psi_{\rm vac}=
\xi_{\rm vac}=0$, and we thus need  only work
to linear order in the variables. From 
Ref.~\cite{gi} the normal mode equations are
\begin{mathletters}%
\label{lineq}
\begin{eqnarray}
\label{giSigmaLin} 
 \Biggl(\partial_\mu\partial^\mu + 4\lambda \Biggr)\,h  &=&0
\\
\label{giRhoLin} 
 \Biggl(\partial_\mu\partial^\mu + \frac{1}{2}  +
 \frac{2}{r^2}\Biggr)\,y &=& 0
\\
\label{giPsiLin} 
 \partial^\mu\left\{ \frac{\partial_\mu \psi -  \epsilon_{\mu\nu}
 \partial^\nu \xi}{1 + \frac{1}{4}r^2}\right\} + \frac{2}{r^2} \,
 \psi &=& 0
\\
\label{giXiLin}
 \partial^\mu\left\{ \frac{ \frac{1}{4} r^2 \partial_\mu \xi +
 \epsilon_{\mu\nu}\partial^\nu \psi }{1 + \frac{1}{4}r^2 }
 \right\} + \frac{1}{2} \xi &=& 0 \ .
\end{eqnarray}
\end{mathletters}%
Equation (\ref{giSigmaLin}) corresponds to a 
pure Higgs excitation characterized by mass 
$M_H=2\sqrt{\lambda}$, while 
(\ref{giRhoLin})-(\ref{giXiLin}) correspond to
three gauge modes of mass  
$M_W = 1/\sqrt{2}$.\footnote[1]{Upon restoring 
the factors of $g$ and the Higgs vacuum
expectation value $v$, these masses take the standard
form $M_H=\sqrt{2\lambda}\, v$ and $M_W = 
(1/2)g\, v$.}

Note that there are four types of normal modes. 
The first two are easily obtained by solving the 
independent equations  (\ref{giSigmaLin}) and 
(\ref{giRhoLin}), while the last two can be 
found by solving the coupled equations 
(\ref{giPsiLin}) and (\ref{giXiLin}) involving 
$\psi$ and $\xi$. A solution in the linear 
regime can then be expanded as a combination 
of these four modes and the amplitudes $a_{k n}$ 
extracted, where $k=1,2,3,4$ specifies the mode 
type. The Higgs and gauge particle numbers are 
defined by 
\begin{eqnarray}
  \nu_{\rm higgs} &=& \sum_{n=1}^{N_{\rm mode}}
  |a_{1n}|^2  
\\
  \nu_{\rm gauge} &=& \sum_{n=1}^{N_{\rm mode}}
  \left\{ |a_{2n}|^2 +
  |a_{3n}|^2 +|a_{4n}|^2 \right\} \ ,
\end{eqnarray}
with total particle number given by
\begin{eqnarray}
  \nu = \nu_{\rm higgs} + \nu_{\rm gauge} \ .
\end{eqnarray}
To avoid counting lattice artifacts we take 
the ultraviolet cutoff on the mode sums to 
be $N_{\rm mode} \sim N/5$ to $N/10$. 

\begin{figure}
\centerline{
\epsfxsize=100mm
\epsfbox{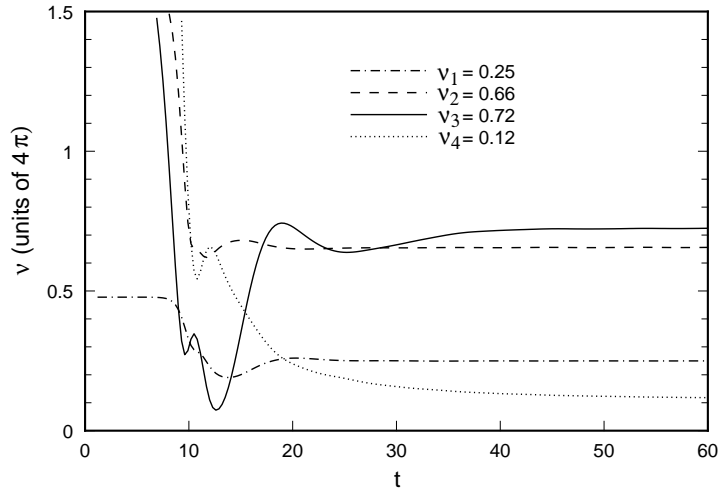}
}
\caption{\tenrm
Decay of a small perturbation about the sphaleron:
behavior of the particle number in the four modes  
as function of time for lattice parameters $N=2239$, 
$\Delta r = 0.04$ and $N_{\rm mode}=200$ with
$\lambda=0.1$. The physical particle numbers are 
obtained by multiplying the asymptotic values in the 
graph by $4\pi/g^2 \sim 30$, which gives 
$N_{\rm higgs}\sim8$ and $N_{\rm gauge}\sim45$, 
for a total physical particle number of $N_{\rm phys}\sim53$.
}
\end{figure}

Space does not permit a detailed exposition
of this procedure, and one should consult 
Ref.~\cite{rs} for full details. Here we must be 
content with Fig.~6, which displays the behavior 
of the particle number in the four normal modes
as a function of time. As an initial
state we chose a typical perturbation about the 
sphaleron as described in the previous section,
and we see that its evolution quickly 
linearizes and settles down into a definite 
asymptotic particle number.

\subsection{Stochastic Sampling of Initial  Configurations}

Recall that our computational strategy consists in 
evolving a configuration near the top of the sphaleron 
barrier until it linearizes, at which point the particle 
number is extracted and the time reversed 
solution is then used to generate the topology changing 
process of interest. We can regard the energy 
$\epsilon$ and the asymptotic particle number $\nu$ 
as functions of the parameters $c_n$ that
specify the initial configuration, and by varying
these coefficients we would like to explore 
the  $\epsilon$-$\nu$ plane and attempt to 
map the region of topology change. In
particular, for a given energy $\epsilon$, 
we would like to find the minimum allowed 
particle number $\nu_{\rm min}(\epsilon)$
consistent with a change of topology. If 
this number can be made arbitrarily small,
this would be a strong indication that baryon
number violation would be observable
in a two-particle collision. 

By randomly exploring the initial configuration 
space, parameterized by the coefficients $c_n$, 
we would stand little chance of making headway. 
Instead, we shall employ stochastic sampling 
techniques, which are ideal for tackling
this type of multi-dimensional minimization.
Our procedure will be to generate initial
configurations  weighted by  $W = \exp(-F)$ 
with $F=\beta \, \epsilon + \mu \, \nu$,
and by adjusting the parameters $\beta$
and $\mu$ we can explore selected
regions in the $\epsilon$-$\nu$ plane.
In particular, by increasing $\mu$ we 
can drive the system to lower and lower 
values of $\nu$ for a given $\epsilon$.
In our numerical work we typically take
$\beta$ between 50 and 1000 while
$\nu$ ranges between 1000 to 20000.

To generate the desired distributions
we have used a Metropolis Monte-Carlo
algorithm. Starting from a definite 
configuration parameterized by 
$c_n$, we perform an upgrade to 
\hbox{$c_n \to c'_n = c_n +\Delta c_n$} 
where $\Delta c_n$ is Gaussian 
distributed with a mean of about 0.0008. 
We evolve the updated configuration
until it linearizes and then calculate
$\Delta F=\beta \, \Delta \epsilon 
+ \mu \, \Delta \nu$. If the topology
of the physically relevant time reversed
solution does not change, then we
discard the updated configuration.
Otherwise we accept it with conditional 
probability $p={\rm Min}[1,\exp(-\Delta F)]$, 
which is equivalent to always accepting 
configurations that decrease $F$ while 
accepting those that increase $F$  with 
conditional probability  $\exp(-\Delta F)$.

\begin{figure}
\centerline{
\epsfxsize=100mm
\epsfbox{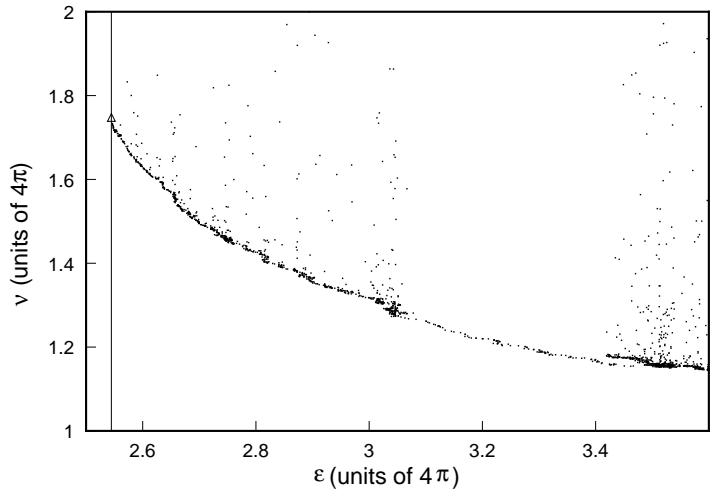}
}
\caption{\tenrm
Monte Carlo results with lattice parameters of $N=2239$,
$\Delta r = 0.04$ (giving L=89.56), $N_{\rm mode}=200$
and $N_{\rm sph}=50$, and with a Higgs self-coupling of
$\lambda=0.1$. The solid line marks the sphaleron
energy $\epsilon_{\rm sph}= 4\pi (2.5426)$, below which
no topology changing process can lie. The triangle represents
the configuration from which we seeded our Monte Carlo
search. To obtain quantities in physical units, multiply the
numbers along the axes by $4\pi/g^2 \sim 30$. The energy axis
extends from about $10 {~\rm TeV}$ to $15 {~\rm TeV}$, while the
particle number axis ranges from about $30$ particles to $60$.
}
\end{figure}

We are now in a position to present our numerical
results. Fig.~7 represents 300 CPU hours 
and involves
30000 solutions (of which only 3000 are shown) 
obtained on the CM-5, a 64 node parallel 
supercomputer.  We have chosen the lattice
parameters \hbox{$N=2239$}, \hbox{$\Delta r=0.04$}, 
with ultraviolet cutoffs determined by 
\hbox{$N_{\rm sph}=50$} and 
\hbox{$N_{\rm mode}=200$}.
The Higgs self-coupling was taken to be
$\lambda=0.1$, which corresponds to a
Higgs mass of $M_H=72 \, {\rm GeV}$.

We have managed to produce a marked decrease
of about 40\% in the minimum particle number 
$\nu_{\rm min}(\epsilon)$, which is approximated
by the lower boundary in the Fig.~7. Nowhere,
however, in the explored energy range does $\nu$
drop below $4\pi$, or in physical units the
incident particle number $N \ge 30$ for energy 
$E \le 15 \, {\rm TeV}$ (the outgoing particle
number tends to be about 50 to 100). This is 
a far cry from two incoming particles which would 
be necessary to argue that baryon number becomes 
unsuppressed in high energy collisions. 

The complex nature of the solution space can
be illustrated by the break in population
density between $\epsilon/4\pi \sim 3 $
and $\epsilon/4\pi \sim 3.4$. In our first 
extended search we did not check whether 
topology change actually occurred, trusting
that the time reversed solutions would continue over
the sphaleron barrier. However, we later found 
an entire region between  $\epsilon/4\pi \sim 3 $
and $\epsilon/4\pi \sim 3.4$ in which the
solutions never left the original topological
sector. We excluded these points and
restarted our search procedure near
$\epsilon/4\pi \sim 3$. A small discontinuity
in the lower boundary with slightly lower particle 
number was produced, but we have still
managed to approximate $\nu_{\rm min}(\epsilon)$
remarkably well. 

We can extract more information from the 
system by investigating the asymptotic
spectral distribution  $|a_{kn}|^2$ as 
a function of mode number $n$. Before 
we started the search, our seed  configuration
(represented by the triangle in Fig.~7)
linearized into a distribution that was 
heavily peaked about a small mode
number $n_{\rm pk} \sim 50$ (with $\Delta n 
\sim 50 $), corresponding to a frequency of
$\omega_{\rm pk} \sim \pi n_{\rm pk}/L \sim 0.1$. 
After the search the solutions underwent 
a dramatic mode redistribution. The amplitudes 
$|a_{kn}|^2$ of the linear regime peaked at 
higher mode number, $n_{\rm pk} \sim 75-100$,
with a much broader distribution ($\Delta n 
\sim 200$). Clearly our search procedure is 
very efficient in redistributing the mode 
population density.

While $\nu$ remains large throughout 
the energy range we have explored, it 
is interesting to note that $\nu_{\rm min}
(\epsilon)$ maintains a slow but steady 
decrease with no sign of leveling off. 
To obtain an indication of the possible
behavior of $\nu_{\rm min}(\epsilon)$
at higher energies, we performed fits
to our data using functional forms which
incorporate expected analytical properties 
of the boundary of the domain of topology
changing solutions. The fits gave a particle 
number $N=2$ at energies in the range of 
$100 \, {\rm TeV}$  to $450 \, {\rm TeV}$. 
Of course we must explore higher energies 
before drawing define conclusions, but 
this is at least suggestive that particle 
number might at some point become small. 

\section{Conclusions}
\hspace{1cm}

We have reviewed in some 
detail the semiclassical method proposed by 
Rubakov, Son, and Tinyakov (RST) for bounding 
the exponential behavior of the two-particle 
baryon number violating cross section in
the standard model. There are two distinct 
regimes, one in which the solutions that 
saturate the functional integral are real, 
corresponding to classically allowed processes 
above the sphaleron barrier, and another
regime in which the saddle-point solutions
are complex, representing Minkowski evolution
followed by Euclidean tunneling under a
barrier. In both cases, a topology changing
solution of small incident particle number
would be a signal that two-particle baryon 
number violating rates would be observable 
in high energy scattering experiments.

Finding the aforementioned saddle-point 
solutions is a formidable numerical task,
but we have nonetheless made considerable 
progress, and in these lectures we have
presented some of our initial computational
results. In the Minkowski regime above the 
sphaleron barrier, we evolve nonlinear 
configurations at the moment of topology 
change until the system linearizes, at which 
point the asymptotic particle number can 
be extracted. The time reversed solutions, 
which have known incident particle numbers, 
will typically undergo topology change, and 
our computational strategy is to stochastically 
search the space of such topology changing 
solutions weighted for small incoming particle 
number. We have found that our numerical algorithm 
is extremely efficient in sifting configurations 
of smaller and smaller incident particle number.
Starting with a generic perturbation of the
sphaleron, which decayed into about 50 particles, 
we have managed to lower the particle number by 
40\%, to approximately 30 particles, while still 
maintaining topology change.  Even though this 
number is rather large, we have only covered 
a narrow energy range, from $10~{\rm TeV}$ 
to $15~{\rm TeV}$, but still it is noteworthy that 
over this domain the data show a slow but steady 
decrease in incident particle number as the energy 
increases. In an effort to increase the rate
at which we can collect numerical data, we have
continued our search on a lattice with half the
number of sites. This new lattice is sufficiently
dense to ensure adequate linearization, but with
much less CPU time, and we soon hope to have 
results well beyond the energy range we have
explored to date. 

Computational methods for finding the saddle-points 
in the classically forbidden regime are more involved 
than the above stochastic procedure. These solutions,
which can become complex along the Euclidean-time
axis, satisfy rather complicated boundary conditions, 
and we are still developing a procedure robust enough 
to find the general RST saddle-point. The periodic 
instanton of Ref.~\cite{krt} is a special case of
these tunneling saddle-points, and it has the advantage 
that it remains real along the entire Euclidean axis 
and satisfies rather simple turning-point boundary 
conditions. In these lectures we have been content 
with presenting a computational procedure to solve 
for the periodic instanton based on conjugate gradient 
minimization, and we have presented a typical numerical 
solution.

\vfill\eject

{\bf Acknowledgments}

C.R. was supported in part under DOE grant
DE-FG02-91ER40676. R.S. was supported in
part under DOE grant DE-FG03-96ER40956 
and NSF grant ASC-940031. We wish to thank
V.~Rubakov for very interesting conversations 
which stimulated the investigation described 
here, and P.~Arnold, A.~Cohen, K.~Rajagopal,
D.~Son, and P.~Tinyakov for valuable discussions.

\end{document}